%% file: preprint3.tex
\documentclass[reprint,showpacs,preprintnumbers,amsmath,amssymb,aps,pra,nofootinbib,floatfix]{revtex4-1}
\usepackage{bbold}
\usepackage{hyperref}

\usepackage{graphicx}

\usepackage{color}
\def\comment#1{{#1}}

\newcommand{\appsection}[1]{\section{\uppercase{#1}}}

\begin{document}

\title{Circumventing Heisenberg's uncertainty principle in atom interferometry tests
of the equivalence principle}

\author{Albert Roura}
\affiliation{Institut f\"ur Quantenphysik, Universit\"at Ulm, Albert-Einstein-Allee 11,
89081 Ulm, Germany}


\date{\today}
\begin{abstract}
Atom interferometry tests of universality of free fall based on the differential measurement of two different atomic species provide a useful complement to those based on macroscopic masses. However, when striving for the highest possible sensitivities, gravity gradients pose a serious challenge.
Indeed, the relative initial position and velocity for the two species need to be controlled with extremely high accuracy, which can be rather demanding in practice and whose verification may require rather long integration times.
Furthermore, in highly sensitive configurations gravity gradients lead to a drastic loss of contrast. These difficulties can be mitigated by employing wave packets with narrower position and momentum widths, but this is ultimately limited by Heisenberg's uncertainty principle.
We present a novel scheme that simultaneously overcomes the loss of contrast and the initial co-location problem. In doing so, it circumvents the fundamental limitations due to Heisenberg's uncertainty principle and eases the experimental realization by relaxing the requirements on initial co-location by several orders of magnitude.
\end{abstract}


\maketitle

\section{Introduction}
\label{sec:introduction}

The equivalence principle is a cornerstone of general relativity and Einstein's key inspirational principle in his quest for a relativistic theory of gravitational phenomena. Experiments searching for small violations of the principle are being pursued in earnest \cite{will14} since they could provide evidence for violations of Loretnz invariance \cite{kostelecky11a} or for dilaton models inspired in string theory \cite{damour10}, and they could offer invaluable hints of a long sought underlying fundamental theory for gravitation and particle physics.
A central aspect which has been tested \comment{to high precision}
is the universality of free fall (UFF) for test masses. Indeed, torsion balance experiments have reached sensitivities at the $10^{-13}\, g$ level \cite{schlamminger08} and it is hoped that this can be improved two orders of magnitude in a forthcoming satellite mission \cite{touboul12}.

An interesting alternative that has been receiving increasing attention in recent years is to perform tests of UFF with quantum systems and, more specifically, using atom interferometry. Instead of macroscopic test masses this kind of experiments compare the gravitational acceleration experienced by atoms of different atomic species \cite{fray04,bonnin13,schlippert14,zhou15}. They offer a valuable complement to traditional tests with macroscopic objects because a wide range of new elements with rather different properties can be employed, so that better bounds on models parameterizing violations of the equivalence principle can be achieved even with lower sensitivities to differential accelerations \cite{hohensee13b,schlippert14}.
Furthermore, given the different kind of systematics involved, they could help to gain confidence in eventual evidence for violations of UFF.

\begin{figure}[h]
\begin{center}
\includegraphics[width=2.8cm]{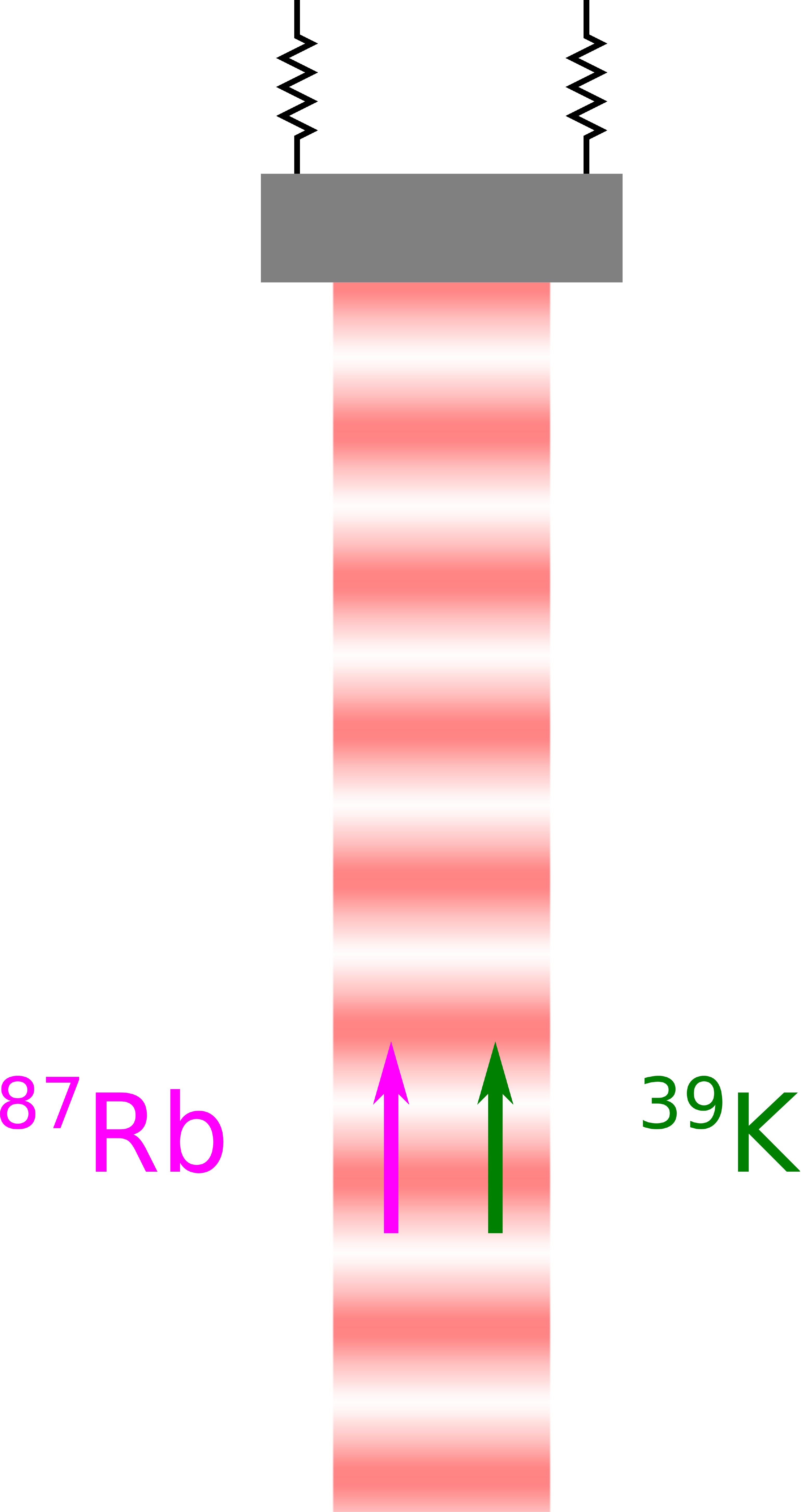}
\end{center}
\caption{Sketch of an atom interferometry set-up for differential acceleration measurements of two different atomic species. The various laser beams driving the diffraction processes for both species share a common retro-reflection mirror so that vibration noise is highly suppressed in the differential phase-shift measurement.}
\label{fig:AI_sketch}
\end{figure}

By using neutral atoms prepared in magnetically insensitive states and an appropriate shielding of the interferometry region, one can greatly suppress the effect of spurious forces acting on the atoms, which constitute excellent inertial references \cite{borde89,kasevich91,peters01}.
State of the art gravimeters based on atom interferometry can reach a precision of the order of $10^{-9}\, g$ in one second \cite{hu13} and are mainly limited by the vibrations of the retro-reflecting mirror. When performing simultaneous differential interferometry measurements for both species and sharing the retro-reflecting mirror (as sketched in Fig.~\ref{fig:AI_sketch}), common-mode rejection techniques can be exploited to suppress the effects of vibration noise and enabling higher sensitivities for the measurement of differential accelerations \cite{fixler07,hogan08,varoquaux09,bonnin13,rosi14}.
Thus, although tests of UFF based on atom interferometry have reached sensitivities up to $10^{-8}\, g$ so far, there are already plans for future space missions that aim for sensitivities of $10^{-15}\, g$ \cite{aguilera14} by exploiting the longer interferometer times available in space and the fact that the sensitivity scales quadratically with the interferometer time.


As shown below, however, when targeting sufficiently high sensitivities, gravity gradients become a great challenge for this kind of experiments: they lead to a drastic loss of contrast in the interference signal and impose the need to control the initial co-location  of the two atomic species (i.e.\ their relative position and velocity) with very high accuracy.
In this paper we will present a novel scheme that simultaneously overcomes both difficulties.

\section{Challenges due to gravity gradients in tests of universality of free fall}
\label{sec:challenges}


In order to analyze the effects of gravity gradients, we will make use of a convenient description of the state evolution in a light-pulse AI developed in Ref.~\cite{roura14} and summarized in \comment{Appendix~\ref{sec:state_evolution}.} 
Within this approach the evolution of the interfering wave packets along each branch of the interferometer is described in terms of centered states $|\psi_\text{c} (t) \rangle$ that characterize their expansion and shape evolution, as well as displacement operators that account for their motion and whose argument $\boldsymbol{\chi} (t) = \big(\boldsymbol{\mathcal{R}}(t), \boldsymbol{\mathcal{P}}(t) \big)^\text{T}$ corresponds to the central position and momentum of the wave packet, which are given by classical phase-space trajectories including the kicks from the laser pulses.
The state at the exit port~I (analogous results hold for the exit port~II) takes then the form
\begin{equation}
|\psi_\text{I} (t)\rangle =
\frac{1}{2} \Big[ e^{i \Phi_1} \hat{\mathcal{D}}(\boldsymbol{\chi}_1)  |\psi_\text{c} (t) \rangle
+ e^{i \Phi_2} \hat{\mathcal{D}}(\boldsymbol{\chi}_2)  |\psi_\text{c} (t) \rangle \Big]
\label{eq:port_I},
\end{equation}
and gives rise to the following oscillations in the fraction of atoms detected in that port as a function of the phase shift $\delta\phi$ between the interferometer branches:
\begin{equation}
\frac{N_\text{I}}{N_\text{I} + N_\text{II}}
= \big\langle \psi_\text{I} (t) \big| \psi_\text{I} (t) \big\rangle
= \frac{1}{2} \big(1 + C \cos \delta\phi \big)
\label{eq:fraction_I}.
\end{equation}
The positive quantity $C$, known as the contrast, characterizes the amplitude of these oscillations and is given by
\begin{equation}
C = \Big| \big\langle \psi_\text{c} (t) \big| \hat{\mathcal{D}}(\delta\boldsymbol{\chi})
\big| \psi_\text{c} (t) \big\rangle \Big| \leq 1
\label{eq:contrast},
\end{equation}
which takes its maximum value when the relative displacement $\delta\boldsymbol{\chi} = \boldsymbol{\chi}_2 - \boldsymbol{\chi}_1$ between the interfering wave packets vanishes.
The full result for the phase shift is provided in \comment{Appendix~\ref{sec:state_evolution}}, but for the present discussion it is sufficient to focus on the role of uniform forces (including inertial ones) and the dependence, caused by gravity gradients, on the central position and velocity ($\mathbf{r}_0$ and $\mathbf{v}_0$) of the initial wave packet:
\begin{equation}
\delta\phi = \mathbf{k}_\text{eff}^\text{T}\, (\mathbf{g} + \mathbf{a}')\, T^2
+ \mathbf{k}_\text{eff}^\text{T}\, \big(\Gamma\, T^2\big) \,
(\mathbf{r}_0 + \mathbf{v}_0 T)
+ \ldots
\label{eq:phase_shift0},
\end{equation}
where for simplicity we have assumed time-independent accelerations and gravity gradients and kept only the contributions to lowest order in $(\Gamma\, T^2)$. We have separated the acceleration $\mathbf{g}$ caused by the gravitational field from the acceleration $\mathbf{a}'$ caused by any other forces including inertial ones and accounting also for the vibrations of the retro-reflecting mirror. On the other hand, the gravity gradient tensor $\Gamma$, defined in Eq.~\eqref{eq:Gamma}, characterizes the deviations from a uniform gravitational field.

\subsection{Initial co-location}
\label{sec:challenges_co-location}

When performing a test of the UFF based on a simultaneous differential measurement with two different atomic species (labeled here $A$ and $B$) the relevant quantity is the phase-shift difference
\begin{align}
\delta\phi^A - \delta\phi^B \approx&\
\mathbf{k}_\text{eff}^\text{T}\, (\mathbf{g}_A - \mathbf{g}_B)\, T^2
+ \mathbf{k}_\text{eff}^\text{T}\, \big(\Gamma\, T^2\big) \, (\mathbf{r}_0^A - \mathbf{r}_0^B)
\nonumber \\
&+\mathbf{k}_\text{eff}^\text{T}\,\big(\Gamma\, T^2\big)\,(\mathbf{v}_0^A - \mathbf{v}_0^B)\,T
\label{eq:co-location}.
\end{align}
Here we have assumed that $\mathbf{k}_\text{eff}$ and $T$ are the same for both species and neglected any contributions to the residual accelerations $\mathbf{a}'$ which are not common to both species. We have made these simplifying assumptions to make the presentation here less cumbersome and to highlight the essential points, but such restrictions are lifted in the remaining sections, so that the strategies and results presented there are more generally applicable.  

As seen from Eq.~\eqref{eq:co-location}, in the presence of gravity gradients small differences in the central position (and velocity) of the initial wave packets for the two atomic species can mimic the effects of a violation of UFF.
In principle preparing wave packets with very well defined central position and momentum does not suffer from limitations associated with Heisenberg's uncertainty principle, which only affects their position and momentum widths. However, the required degree of control on those quantities is rather demanding in practice: for example, testing the UFF at the level of
$10^{-15} g$
entails controlling the relative initial position and velocity with accuracies better than a few nm and several hundred pm/s respectively.
In fact, this systematic effect, often known in this context as the \emph{initial co-location} problem, is one of the biggest challenges faced by this kind of experiments. 
Furthermore, verifying that the stringent requirements on initial co-location are fulfilled by measuring the relative position and velocity of the two species under the same experimental conditions as in the differential interferometry measurements (otherwise one could introduce additional systematic effects) is itself limited by Heisenberg's uncertainty principle, which implies
\begin{equation}
n\, N\, \sigma_x\, \sigma_p \geq \hbar / 2 \,
\label{eq:HUP},
\end{equation}
where $\sigma_x$ and $\sigma_p$ are the precisions for the measurement of the central position and momentum that can be achieved after a given integration time, $N$ is the number of atoms in each atom cloud and $n$ is the number of experimental runs, given by the integration time multiplied by the repetition rate.
Thus, unless the number of atoms in each atom cloud is rather high, the required integration time to achieve the desired accuracy in these checks of systematics may be comparable to or even exceed the entire lifetime of the mission, which has been raised as an objection against the use of atoms (rather than macroscopic test masses) as inertial references for high-precision tests of UFF \cite{nobili15}.

\subsection{Loss of contrast}
\label{sec:challenges_contrast}

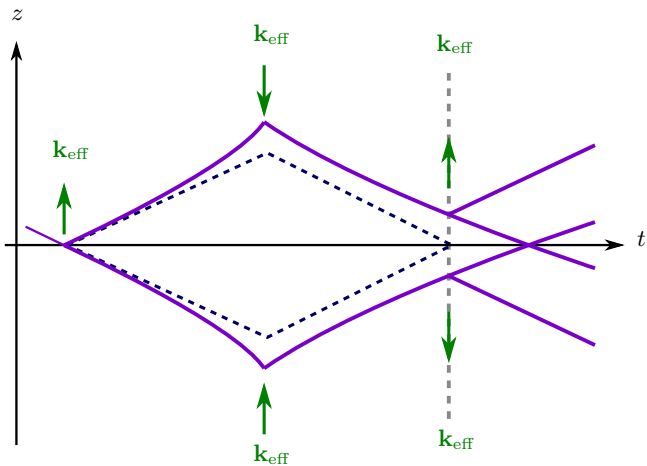
\begin{figure}[h]
\begin{center}
\def\svgwidth{8.5cm}
\input{figure2_pdf.tex}
\end{center}
\caption{Central trajectories for a Mach-Zehnder interferometer in the presence of gravity gradients as seen in a suitable freely falling frame. The tidal forces tend to open up the trajectories compared to the case without gravity gradients (dashed lines).
The momentum transfer from the laser pulses are also indicated for the different branches.
}
\label{fig:gg_trajectories}
\end{figure}

In order to understand qualitatively the motion of the atomic wave packets along the branches of the interferometer in the presence of a gravity gradient, it is useful to consider a freely falling frame where the central trajectories for the two branches after the first beam-splitter pulse are symmetric with respect to the spatial origin of coordinates at $z=0$, as depicted in Fig.~\ref{fig:gg_trajectories}. It then becomes clear that the tidal forces associated with the gravity gradient tend to open up these trajectories and give rise to \comment{an \emph{open interferometer} with} non-vanishing relative position and momentum displacements at the exit ports. They are quantitatively obtained in \comment{Appendix~\ref{sec:state_evolution}} and are given by
\begin{equation}
\delta \boldsymbol{\mathcal{R}} = \left(\Gamma\, T^2 \right) \mathbf{v}_\text{rec} \, T \, ,
\quad
\delta \boldsymbol{\mathcal{P}} = (\Gamma\, T^2) \, m\, \mathbf{v}_\text{rec}
\label{eq:deltaR_deltaP} \, .
\end{equation}
The existence of a non-vanishing relative displacement $\delta\boldsymbol{\chi} \neq 0$ between the interfering wave packets, which no longer exhibit full quantum overlap, gives rise to a loss of contrast, as reflected by Eq.~\eqref{eq:contrast}.

In terms of the Wigner function \cite{hillary84}, defined in Eq.~\eqref{eq:wigner_def}, the expression for the contrast takes the following suggestive form, which is also valid for mixed states~\cite{roura14}:
\begin{equation}
C = \left | \, \int d^3x \int d^3p \, W(\mathbf{x},\mathbf{p};t) \,
e^{-\frac{i}{\hbar} \delta \boldsymbol{\chi}^\text{T} J \, \boldsymbol{\xi}} \, \right |
\label{eq:contrast_wigner} ,
\end{equation}
where we introduced the phase-space vector $\boldsymbol{\xi} = (\mathbf{x}, \mathbf{p})^\text{T}$ and the symplectic form $J$ defined in Eq.~\eqref{eq:symplectic_form}.
The loss of contrast can be understood as a consequence of the oscillatory factor ``washing out'' the result of the integral. It otherwise tends to unity when the oscillations are negligible, as dictated by the normalization of the Wigner function.
The example of Gaussian states, whose Wigner function is given by Eq.~\eqref{eq:wigner_gaussian}, is particularly illustrative. Substituting into Eq.~\eqref{eq:contrast_wigner}, one obtains the following result for the contrast:
\begin{equation}
C = \exp \Big[ -\frac{1}{2 \hbar^2}\, \delta \boldsymbol{\chi}_0^\text{T} J^\text{T} \Sigma \,
J \, \delta \boldsymbol{\chi}_0 \Big]
\label{eq:contrast_gaussian},
\end{equation}
where $\Sigma$ is the covariance matrix of the initial state, detailed in Eq.~\eqref{eq:covariance}.
In turn, $\delta \boldsymbol{\chi}_0 \equiv \mathcal{T}^{-1} (t,t_0) \, \delta \boldsymbol{\chi}$ is related to $\delta \boldsymbol{\chi}$ through the transition matrix $\mathcal{T} (t,t_0)$ as explained in Appendix~\ref{sec:classical_trajectories} and well approximated by Eqs.~\eqref{eq:deltaR_free}-\eqref{eq:deltaP_free} for gravity gradients.
It is clear from Eq.~\eqref{eq:contrast_gaussian}
that the loss of contrast caused by gravity gradients can be reduced by simultaneously considering smaller position and momentum spreads, $\Sigma_{xx}$ and $\Sigma_{pp}$. 
In fact, such a conclusion is not specific of Gaussian states and holds in general. Indeed, as the size of the main support of the Wigner function decreases, the wash-out effect from the oscillatory factor becomes less and less important.

Simultaneously decreasing $\Sigma_{xx}$ and $\Sigma_{pp}$ is, however, ultimately limited by Heisenberg's uncertainty principle. Moreover, achieving a sufficiently narrow momentum distribution can sometimes be rather demanding in practice even before the limit due to Heisenberg's uncertainty principle is reached. In order to alleviate these difficulties, an easily implementable mitigation strategy based on a small adjustment $\delta T / T \sim (\Gamma_{zz}\, T^2)$ of the timing for the last pulse was proposed in Ref.~\cite{roura14}. The key idea is that with a suitable choice of $\delta T$  one can change $\delta \boldsymbol{\mathcal{R}}$ (while keeping $\delta \boldsymbol{\mathcal{P}}$ essentially unchanged) so that the phase-space vector $(J\, \delta \boldsymbol{\chi})$ becomes aligned with the Wigner function and the wash-out effect of the oscillatory factor in Eq.~\eqref{eq:contrast_wigner} is minimized. (This is actually reminiscent of the use of squeezed states to beat the standard quantum limit in optical interferometers; rather than squeezing the state, here the observable is modified to achieve a similar effect.)
The strategy, which is briefly summarized in \comment{Appendix~\ref{sec:contrast_wigner_gaussian}}, was shown to be very effective for parameter ranges like those considered, for example, for the STE-QUEST mission \cite{aguilera14}. Nevertheless, it would eventually face significant limitations in future plans for further increasing by several orders of magnitude the sensitivity of UFF tests based on atom interferometry \cite{dimopoulos07}. Furthermore, it is somewhat less effective when applied to thermal clouds (even for ultracold atoms close to quantum degeneracy but with a negligible condensate fraction) rather than Bose-Einstein condensates%
\footnote{\comment{These limitations also apply to an alternative approach based on extracting the phase shift from the spatial location of the maxima within the fringe pattern that arises in the density profile at the exit ports of an open interferometer \cite{muentinga13,sugarbaker13}.}}.
In the next section we present a novel scheme which is not afflicted by these shortcomings and simultaneously overcomes both the initial co-location problem and the loss of contrast.

\section{Simultaneously overcoming loss of contrast and the initial co-location problem}
\label{sec:co-location}

\begin{figure}[h]
\begin{center}
\def\svgwidth{8.5cm}
\input{figure3_pdf.tex}
\end{center}
\caption{Central trajectories for the same situation depicted in Fig.~\ref{fig:gg_trajectories} but with a suitable adjustment of the momentum transfer from the second laser pulse so that a closed interferometer, with vanishing relative displacement between the interfering wave packets in each port, is recovered.}
\label{fig:mitigation_trajectories}
\end{figure}
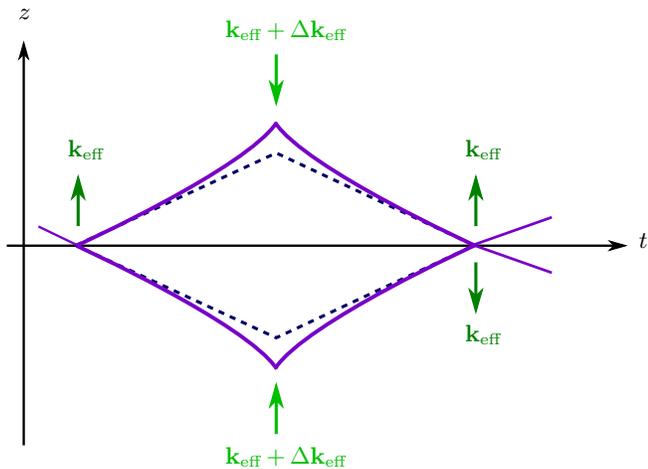

As shown by Eq.~\eqref{eq:phase_shift3}, the contributions to the phase shift $\delta\phi$ that depend on the initial values of the central position and momentum, characterized by the phase-space vector $\boldsymbol{\chi}_0 = (\mathbf{r}_0, m \mathbf{v}_0)^\text{T}$, can be written in the following revealing form:
\begin{equation}
\delta\phi = - \frac{1}{\hbar}\, \delta \boldsymbol{\chi}^\text{T}(t)\, J \, \mathcal{T}(t,t_0) \, \boldsymbol{\chi}_0 + \ldots
\label{eq:phase_shift_initial},
\end{equation}
where $\mathcal{T}(t,t_0)$ is the transition matrix defined right after Eq.~\eqref{eq:full_solution}.
The next insight is then to realize that through a suitable adjustment of the laser wavelength for the second pulse, one can have a vanishing final displacement $\delta\boldsymbol{\chi} = 0$.
Indeed, this can be achieved by changing the effective momentum transfer associated with the second pulse to $\hbar\, \big( \mathbf{k}_\text{eff} + \Delta \mathbf{k}_\text{eff} \big)$ with 
\begin{equation}
\Delta \mathbf{k}_\text{eff} = \big( \Gamma\, T^2 / 2 \big)\, \mathbf{k}_\text{eff}
\label{eq:Delta_k},
\end{equation}
where we neglected corrections of higher order in $\big( \Gamma\, T^2 \big)$ for simplicity [the exact result can be obtained using Eq.~\eqref{eq:transition_exact}]. This can be easily understood in a suitable freely falling frame where the central trajectories for the two branches of the interferometer are symmetric with respect to $z=0$, as shown in Fig.~\ref{fig:mitigation_trajectories}. The momentum transfer associated with the second pulse is chosen so that the trajectory after the pulse corresponds to the time reversal of the trajectory before it. Since the gravity gradient leads to the curvature of the trajectories in the space-time diagram as depicted in Fig.~\ref{fig:mitigation_trajectories}, an increase of the momentum transfer given by Eq.~\eqref{eq:Delta_k} is necessary.

From Eqs.~\eqref{eq:contrast} and \eqref{eq:phase_shift_initial} it is clear that by achieving $\delta\boldsymbol{\chi} = 0$, this scheme simultaneously takes care of the loss of contrast caused by gravity gradients as well as the stringent requirements on the initial co-location of the two atomic species.
Indeed, thanks to the absence of a relative displacement between the interfering wave packets full contrast is recovered.
Furthermore, there is also an intuitive explanation for the solution of the initial co-location problem which is connected with the laser phases.
Because of the additional $\Delta \mathbf{k}_\text{eff}$ for the second pulse, in this new scheme the total momentum transfer to the two branches is unbalanced, i.e.\ one no longer has $\sum_j \delta \mathbf{k}_\text{eff}^{(j)} = 0$, where $\hbar\, \delta \mathbf{k}_\text{eff}^{(j)}$ corresponds to the difference between the momenta transferred to the two branches by the $j$-th pulse. This implies that the total contribution from the laser phases exhibits the following dependence on the initial values of the central position and velocity of the atomic wave packet:
\begin{align}
\delta\phi_\text{laser} \, \to \, &
\sum_j \delta \mathbf{k}_\text{eff}^{(j)} \cdot (\mathbf{r}_0 + \mathbf{v}_0 T) =
-2\, \Delta \mathbf{k}_\text{eff}^\text{T}\, (\mathbf{r}_0 + \mathbf{v}_0 T)  \nonumber \\
& \qquad\qquad\qquad = -\mathbf{k}_\text{eff}^\text{T}\, \big(\Gamma T^2\big) \,
(\mathbf{r}_0 + \mathbf{v}_0 T)
\label{eq:laser_phase} ,
\end{align}
where we made use of the choice of $\Delta \mathbf{k}_\text{eff}$ specified above in the second equality.
The key point is that the dependence on the initial position and velocity of the contribution from the laser phases compensates the effect of gravity gradients because the right-hand side of Eq.~\eqref{eq:laser_phase} exactly cancels the terms depending on the initial conditions in Eq.~\eqref{eq:phase_shift0}.

On the other hand, the fact that $\sum_j \delta \mathbf{k}_\text{eff}^{(j)} \neq 0$ also implies that the phase shift depends on the initial position (and velocity) of the retro-reflecting mirror, which is not known or controlled with very high precision.
More specifically, it gives rise to the following contribution to the phase shift:
\begin{equation}
\delta\phi_\text{laser} \, \to \, -\, 2\, \Delta\mathbf{k}_\text{eff} \cdot \mathbf{r}_\text{mirror} 
= -2\, \mathbf{k}_\text{eff}^\text{T}\, \big(\Gamma\, \mathbf{r}_\text{mirror}\big) \, T^2
\label{eq:mirror_position} ,
\end{equation}
where $\mathbf{r}_\text{mirror}$ corresponds to the position of the retro-reflecting mirror at the time the second pulse is applied. The effect of this contribution can be post-corrected for a known mirror position $\mathbf{r}_\text{mirror}$. However, an uncertainty $\Delta\mathbf{r}_\text{mirror}$ on the position of the mirror leaves a residual contribution to the phase shift after post-correction corresponding to the expression in Eq.~\eqref{eq:mirror_position} but with $\mathbf{r}_\text{mirror}$ replaced by $\Delta\mathbf{r}_\text{mirror}$.
Fortunately, such dependence on $\Delta\mathbf{r}_\text{mirror}$ drops out (to sufficiently high degree) in the differential measurement. Indeed, when performing a differential measurement between species $A$ and $B$, the dependence on the position uncertainty $\Delta\mathbf{r}_\text{mirror}$ reduces to 
\begin{equation}
\delta\phi_\text{laser}^A - \delta\phi_\text{laser}^B \, \to \, -2\, \big( \mathbf{k}_A\, T_A^2
- \mathbf{k}_B\, T_B^2 \big)^\text{T}
\big(\Gamma\, \Delta\mathbf{r}_\text{mirror}\big)
\label{eq:mirror_position_diff} ,
\end{equation}
which can be made small enough in exactly the same way as the contribution to the differential measurement of  residual accelerations common to both species.
For example, given the gravity gradient on Earth' surface, one has $\big| \Gamma\, \Delta\mathbf{r}_\text{mirror} \big| \sim 3\times10^{-13}\, g $ for an uncertainty $| \Delta \mathbf{r}_\text{mirror} | = 1\, \mu\text{m}$ in the position of the mirror. This means, for example, that if one can suppress residual accelerations $a' \lesssim 10^{-12}\, g$ through common-mode rejection, any contribution due to the unknown position of the mirror will be suppressed provided that this can be determined with an uncertainty $| \Delta \mathbf{r}_\text{mirror} | \lesssim 1\, \mu \text{m}$.
In fact, typical plans for high-precision tests of UFF with AIs are designed to reject much higher residual accelerations, so that the requirements on $| \Delta \mathbf{r}_\text{mirror} |$ can be further relaxed by several  orders of magnitude compared to this example.

So far we have assumed perfect knowledge of the gravity gradients, which would allow perfect compensation of their effects (up to the uncertainty in the mirror position) by using the scheme described above with $\Delta \mathbf{k}_\text{eff}$ given by Eq.~\eqref{eq:Delta_k}. In practice, however, the gravity gradient tensor $\Gamma$ employed in Eqs.~\eqref{eq:Delta_k}-\eqref{eq:mirror_position_diff} needs to be determined by simulations of the mass distribution, direct gradiometry measurements or a combination of both.
Since this can only be done with some finite accuracy $\Delta\Gamma$ characterizing the difference between the actual gravity gradients and the tensor $\Gamma$ determined by those means, the final displacement $\delta\boldsymbol{\chi}$ will take some non-vanishing residual value with $\Delta\Gamma$ appearing instead of $\Gamma$ in Eqs.~\eqref{eq:deltaR_deltaP}. Consequently, the dependence of the phase shift on the initial conditions appearing in Eq.~\eqref{eq:phase_shift_initial} will not be completely eliminated, but the co-location requirements for the differential measurement will be substantially relaxed by a factor of order $\|\Delta\Gamma\| / \|\Gamma\|$. Hence, determining for instance the gravity gradient tensor $\Gamma$ with a relative accuracy $\|\Delta\Gamma\| / \|\Gamma\| \lesssim 10^{-3}$ leads to a relaxation of the initial co-location requirements by 3 orders of magnitude.

We conclude this section by briefly discussing how feasible it is to implement the required momentum change for the second pulse, which is given by Eq.~\eqref{eq:Delta_k} and amounts to a relative frequency change $\Delta\nu / \nu = (\Gamma_{zz}\, T^2)$ if the gravity gradient is \comment{aligned} with $\mathbf{k}_\text{eff}$. Given Earth's gravity gradient $\Gamma_{zz} \approx 3 \times 10^{-6}\, \text{s}^{-2}$ and a moderate interferometer time $2\,T = 2\,\text{s}$, this corresponds to a frequency change $\Delta\nu \approx 1\, \text{GHz}$, which can be easily implemented with acousto-optical modulators (AOMs). This single-photon frequency change will be the same even if one has a large momentum transfer through higher-order Bragg diffraction, a sequence of multiple $\pi$ pulses or a combination of both. On the other hand, for a longer interferometer time $2\,T = 10\,\text{s}$ a single-photon frequency change $\Delta\nu \approx 25\, \text{GHz}$ would be necessary. Such a frequency change requires a more sophisticated set-up \comment{(e.g.\ two phase-locked lasers)}. Moreover, it gives rise to a substantially larger detuning of the single-photon transition, which requires a higher laser intensity in order to have a comparable Rabi frequency.
(The detuning could be approximately reduced in half by going from red- to blue-detuned transitions when changing from $\mathbf{k}_\text{eff}$ to $\mathbf{k}_\text{eff} + \Delta \mathbf{k}_\text{eff}$.)
Thus, the new scheme seems somewhat easier to implement directly in set-ups where higher sensitivity is achieved through large momentum transfer rather than very long interferometer times.



\section{Conclusion}
\label{sec:conclusion}

Gravity gradients pose a great challenge for high-precision tests of UFF based on atom interferometry. Indeed, for sufficiently large values of the effective momentum transfer or the interferometer time gravity gradients lead to a drastic loss of contrast and impose a serious limitation on the highest sensitivity that can be achieved.
Furthermore, they also imply stringent requirements on the initial co-location of the wave packets for the two species since the effects of a non-uniform gravitational field could otherwise mimic a violation of UFF.
Although there is in principle no fundamental limitation on the precision with which the central position and momentum of the wave packets for both species can be determined, in practice the requirements can be rather challenging. Moreover, the time needed to verify that such requirements are fulfilled can be comparable to the entire mission lifetime.

The situation concerning the loss of contrast and the time needed for verification of the systematics associated with initial co-location can be improved by considering smaller position and momentum widths for the initial state, but this is ultimately limited by Heisenberg's uncertainty principle.
In this paper we have presented a novel scheme that simultaneously overcomes the loss of contrast and the initial co-location problem. In doing so, it circumvents the limitations due to Heisenberg's uncertainty principle on the highest sensitivities that can be achieved and eases the experimental realization by relaxing the requirements on initial co-location by several orders of magnitude.
The key idea is that by slightly changing the wavelength of the second laser pulse, the momentum transfer to the two interferometer branches becomes unbalanced and this implies that the total phase-shift contribution from the laser phases depends on the initial values of the central position and momentum of the atomic wave packets. In fact, with a suitable choice of the change of wavelength this can exactly compensate the analogous contribution caused by the gravity gradients. 
Furthermore, this choice automatically gives rise to a closed interferometer (vanishing relative displacement between the interfering wave packets) with no loss of contrast.

The results and discussions presented here can also be applied directly to experiments performed under microgravity conditions (this would actually correspond to the freely falling frame considered in Figs.~\ref{fig:gg_trajectories} and \ref{fig:mitigation_trajectories}). In that case employing a retro-reflection set-up naturally leads to double diffraction \cite{leveque09,giese13} and \comment{the associated} symmetric interferometers, which have a number of advantages, such as immunity to a number of systematic effects and noise sources (including laser phase noise). Only one detail needs then to be changed in Figs.~\ref{fig:gg_trajectories} and \ref{fig:mitigation_trajectories}: the central velocity of the initial wave packet vanishes and there is an additional third exit port with vanishing central velocity.
Interestingly, the double diffraction scheme can be generalized to experiments performed in a laboratory under normal gravity conditions by adding a third laser frequency per species \cite{malossi10,zhou15}, so that one can still benefit from many of the advantages associated with double diffraction (including also the cancellation of the photon-recoil term quadratic in $\mathbf{k}_\text{eff}$ mentioned in Appendix~\ref{sec:phase_shift}).

Rotations, which were not explicitly considered here, also lead to open interferometers, loss of contrast and dependence of the phase shift on the central velocity of the initial wave packet. These effects can be mitigated by employing satellites with sufficiently small angular velocity, the use of a tip-tilt mirror that \comment{compensates rotations} \cite{hogan08,lan12,dickerson13} or a combination of both. Moreover, by combining it with  the use of such a tip-tilt mirror, the scheme presented here can be extended to the case of non-aligned gravity gradients, where the direction of $\mathbf{k}_\text{eff}$ does not coincide with a principle axis of the tensor $\Gamma$.

Finally, it is worth pointing out that \comment{our method} can also be applied to differential measurements involving two (or more) spatially separated atom interferometers interrogated by common laser beams, a configuration employed for gradiometry measurements \cite{snadden98,rosi15}. Indeed, a scheme completely analogous to that presented here could be  exploited to mitigate the loss of contrast for each single interferometer (for sufficiently high $\mathbf{k}_\text{eff}$ or interferometer time) as well as the additional loss of contrast in the differential measurement due to the coupling of static gravity gradients to initial position and velocity jitter from shot to shot, which can become particularly important when considering long baselines between the interferometers. This would be relevant for gravitational antennas  capable of monitoring very precisely changes in the local gravitational field, such as the MIGA facility \cite{geiger15} currently under construction, and with interesting applications to geophysics and hydrology. It may also be relevant for gravitational antennas with very long baselines which have been proposed for the detection of low-frequency gravitational waves \cite{dimopoulos08b,graham13,hogan15}.



%
%
%
%
%
%
%

\begin{acknowledgments}
This work was supported by the German Space Agency (DLR) with funds provided by the Federal Ministry of Economics and Technology (BMWi) under Grant No.~50WM1556 (QUANTUS IV).
It is a pleasure to thank Wolfgang Zeller, Stephan Kleinert and Wolfgang Schleich for collaboration in related earlier work.
\end{acknowledgments}


\bibliographystyle{apsrev4-1}
\bibliography{literature}


\onecolumngrid
\newpage

\begin{center}
{\Large\bf Supplemental Material}
\end{center}
\vspace{2.0em}
\twocolumngrid


\appendix

\appsection{State evolution in a light-pulse atom interferometer}
\label{sec:state_evolution}

In this Appendix we summarize a convenient formalism developed in Ref.~\cite{roura14} for the description of the state evolution in a light-pulse atom interferometer. The dynamics between laser pulses is governed by the following Hamiltonian operator, which can naturally account (in particular) for uniform gravitational and inertial forces as well as gravity gradients:
\begin{equation}
\hat{H} = \frac{1}{2m} \hat{\mathbf{p}}^\text{T} \hat{\mathbf{p}}
- \frac{m}{2} \hat{\mathbf{x}}^\text{T} \Gamma (t)\, \hat{\mathbf{x}}
- m\, \mathbf{g}^\text{T} (t) \, \hat{\mathbf{x}} + V_0(t)
\label{eq:hamiltonian},
\end{equation}
where the vector $\mathbf{g}$ characterizes the acceleration associated with the uniform forces and the spatially independent potential $V_0(t)$ can incorporate possibly branch-dependent internal energies. In turn, $\Gamma$ is a symmetric tensor corresponding to the Hessian of the gravitational potential $U(\mathbf{x})$:
\begin{equation}
\Gamma_{ij} = -\frac{\partial^2 U}{\partial x^i \partial x^j}
\label{eq:Gamma} \, .
\end{equation}

Let us consider an arbitrary initial state
\begin{equation}
|\psi (t_0)\rangle = \hat{\mathcal{D}} \big( \boldsymbol{\chi}_0  \big) \,
| \psi_\text{c} (t_0) \rangle
\label{eq:initial_state},
\end{equation}
written in terms of a centered state $| \psi_\text{c} (t_0) \rangle$ with vanishing position and momentum expectation values, and a displacement operator whose argument contains the information on the central position and momentum of the wave packet.
The displacement operator $\hat{\mathcal{D}}(\boldsymbol{\chi})$ is defined as
\begin{equation}
\hat{\mathcal{D}}(\boldsymbol{\chi}) = e^{-\frac{i}{\hbar} \boldsymbol{\chi}^\text{T} J \, \hat{\boldsymbol{\xi}}}
\label{eq:displacement} ,
\end{equation}
where we used a vector notation for phase-space quantities, so that $\hat{\boldsymbol{\xi}} = (\hat{\mathbf{x}},\hat{\mathbf{p}})^\text{T}$, and introduced the symplectic form
\begin{equation}
J = \left( \begin{array}{cc} 0 & \mathbb{1} \\ -\mathbb{1} & 0 \end{array} \right)
\label{eq:symplectic_form}.
\end{equation}
It can be shown that the time evolution of the initial state~\eqref{eq:initial_state} along each branch is given by  
\begin{equation}
|\psi (t)\rangle = e^{i \Phi(t)}\, \hat{\mathcal{D}} \big( \boldsymbol{\chi}(t)  \big) \, | \psi_\text{c} (t) \rangle
\label{eq:state_evol},
\end{equation}
where the centered state $| \psi_\text{c} (t) \rangle$ evolves according to the purely quadratic part of the Hamiltonian~\eqref{eq:hamiltonian}.
In the absence of laser pulses the displacement vector
$\boldsymbol{\chi} (t) = \big(\boldsymbol{\mathcal{R}}(t), \boldsymbol{\mathcal{P}}(t) \big)^\text{T}$ corresponds to the classical phase-space trajectories associated with the Hamiltonian~\eqref{eq:hamiltonian} and with initial conditions $\boldsymbol{\chi} (t_0) = \boldsymbol{\chi}_0$.

On the other hand, assuming idealized laser pulses where the effects of pulse duration and dispersion effects are neglected, the action of each pulse on the atom's center-of-mass motion can be represented with the following phase factor and displacement operator:
\begin{equation}
e^{i \varepsilon_j\varphi_j}\, e^{-i |\varepsilon_j| \frac{\pi}{2}} \,
\hat{\mathcal{D}} \big( \boldsymbol{\mathcal{F}}_j \big)
\label{eq:laser_pulse} ,
\end{equation}
where $\varepsilon_j = 0, \pm1$ for each pulse and takes different values depending on the interferometer branch, $\varphi_j$ is a spatially independent phase for the $j$-th pulse and 
\begin{equation}
\boldsymbol{\mathcal{F}}_j = \varepsilon_j  \left( \begin{array}{c} \mathbf{0} \\ \hbar \mathbf{k}_j \end{array} \right)
\label{eq:laser_kick},
\end{equation}
characterizes the momentum kick from that pulse for each branch.
Making use of the composition formula for displacement operators
\begin{equation}
\hat{\mathcal{D}}(\boldsymbol{\chi}) \, \hat{\mathcal{D}}(\boldsymbol{\chi}')
= e^{-\frac{i}{2 \hbar} \boldsymbol{\chi}^\text{T} J \, \boldsymbol{\chi}'} \,
\hat{\mathcal{D}}(\boldsymbol{\chi} + \boldsymbol{\chi}')
\label{eq:composition},
\end{equation}
the operator~\eqref{eq:laser_pulse} for each pulse can be combined with the result for the state evolution between laser pulses mentioned above. Proceeding recursively, one finds that when the pulses are taken into account, the displacement $\boldsymbol{\chi} (t)$ in Eq.~\eqref{eq:state_evol} is still given by the classical trajectories associated with the Hamiltonian~\eqref{eq:hamiltonian}, but including the instantaneous kicks from the various pulses.
Furthermore, the following result is obtained for the phase $\Phi(t)$:
\begin{equation}
\Phi(t) = \varphi \,-\, \frac{1}{2\hbar}\! \int_{t_0}^t \! dt' \Bigg( \Big[ \boldsymbol{\mathcal{F}}_\text{lp}^\text{T}(t') + \boldsymbol{\mathcal{G}}^\text{T}(t') \Big] J\, \boldsymbol{\chi} (t') \,+\, 2V_0 (t') \Bigg)
\label{eq:total_phase},
\end{equation}
with $\varphi = \sum_{j=1}^n \big( \varepsilon_j\, \varphi_j - |\varepsilon_j| \, \pi/2 \big)$ and
\begin{align}
\boldsymbol{\mathcal{F}}_\text{lp} (t) &= \sum_{i=1}^n \delta(t-t_i) \, \boldsymbol{\mathcal{F}}_i 
\label{eq:laser_force}, \\
\boldsymbol{\mathcal{G}}(t) &=
\left( \begin{array}{c} \mathbf{0} \\ m \mathbf{g}(t) \end{array} \right) 
\label{eq:g_force}.
\end{align}
Note that the uniform forces, encoded in $\boldsymbol{\mathcal{G}}(t)$, enter exactly in the same way as the laser kicks but with a continuous time dependence rather than a finite set of instants.
This fact was exploited in Ref.~\cite{roura14} to simplify considerably the derivation of Eq.~\eqref{eq:total_phase} and that of the general phase shift formula in Sec.~\ref{sec:phase_shift} below.

\subsection{Classical trajectories}
\label{sec:classical_trajectories}

As explained above, the phase-space vector $\boldsymbol{\chi} (t)$ corresponding to the trajectory of the center of the wave packet for each branch is a solution of the classical equations of motion associated with the Hamiltonian~\eqref{eq:hamiltonian} together with the additional kicks from the laser pulses. Therefore, it satisfies the following equation:
\begin{equation}
\dot{\boldsymbol{\chi}} (t) - \mathcal{H}(t) \, \boldsymbol{\chi} (t)
= \boldsymbol{\mathcal{G}}(t) + \boldsymbol{\mathcal{F}}_\text{lp} (t)
\label{eq:eom},
\end{equation}
where
\begin{equation}
\mathcal{H}(t) =
\left( \begin{array}{cc} 0 & (1/m) \, \mathbb{1} \\ m \, \Gamma(t) & 0 \end{array} \right) 
\label{eq:eom_op},
\end{equation}
and with $\boldsymbol{\mathcal{F}}_\text{lp}(t)$ and $\boldsymbol{\mathcal{G}}(t)$ given by Eqs.~\eqref{eq:laser_force}-\eqref{eq:g_force}.
The solutions of Eq.~\eqref{eq:eom} are uniquely specified by the initial conditions $\boldsymbol{\chi} (t_0) = \boldsymbol{\chi}_0$ and are given by
\begin{equation}
\boldsymbol{\chi} (t) = \mathcal{T} (t,t_0)\, \boldsymbol{\chi}_0
+ (\mathcal{T}_\text{ret} \cdot \boldsymbol{\mathcal{G}}) (t)
+ (\mathcal{T}_\text{ret} \cdot \boldsymbol{\mathcal{F}}_\text{lp}) (t)
\label{eq:full_solution},
\end{equation}
where the transition matrix $\mathcal{T} (t,t_0)$ satisfies the homogeneous part of Eq.~\eqref{eq:eom} with initial condition $\mathcal{T} (t_0,t_0) = \mathbb{1}$, and we employed the retarded propagator $\mathcal{T}_\text{ret} (t,t') = \mathcal{T} (t,t') \, \theta(t-t')$ and introduced the notation
\begin{equation}
(\mathcal{T}_\text{ret} \cdot \boldsymbol{\mathcal{A}}) (t) \equiv \int^t_{t_0} dt' \, \mathcal{T}_\text{ret} (t,t')\, \boldsymbol{\mathcal{A}} (t')
\label{eq:convolution} .
\end{equation}

For a time-independent gravity gradient tensor the transition matrix can be straightforwardly obtained by exponentiating the matrix $\mathcal{H}(t)$:
\begin{equation}
\mathcal{T} (t,t') =
\left( \begin{array}{cc}
\cosh \left[\gamma (t-t')\right]
& \frac{1}{m\gamma} \sinh \left[\gamma (t-t')\right] \\
m \gamma\, \sinh \left[\gamma (t-t')\right]
& \cosh \left[\gamma (t-t')\right]
\end{array} \right)
\label{eq:transition_exact},
\end{equation}
where we introduced $\gamma \equiv \sqrt{\Gamma}$. In order to calculate explicitly the transition matrix in Eq.~\eqref{eq:transition_exact}, one needs to diagonalize the symmetric tensor $\Gamma$, which is always possible with an orthogonal transformation (a rotation of the coordinate axes). In this new coordinate system the motion along each one of the three axes (known as principal axes) decouples and the dynamics reduces to that of three independent one-dimensional systems.
In typical cases of interest in atom interferometry the condition $|\Gamma_{jj}| T^2 \ll 1$ is amply satisfied for the three principal axes and it is an excellent approximation to consider the following perturbative expansion of Eq.~\eqref{eq:transition_exact} up to linear order in $\Gamma$ (the expansion involves only even powers of $\gamma)$:
\begin{equation}
\mathcal{T} (t,t') \approx
\left( \begin{array}{cc}
\mathbb{1} + \frac{\Gamma}{2} (t-t')^2
& \frac{(t-t')}{m} \Big[\mathbb{1} + \frac{\Gamma}{6} (t-t')^2 \Big] \\
m \, \Gamma \, (t-t')
& \mathbb{1} + \frac{\Gamma}{2} (t-t')^2
\end{array} \right)
\label{eq:transition_linear},
\end{equation}
where we neglected terms of higher order in $\Gamma (t-t')^2$.

Unless the force $m\, \mathbf{g}(t)$ is branch dependent, the first two terms on the right-hand side of Eq.~\eqref{eq:full_solution} are common for the two interferometer branches,
so that only the third term contributes to the relative displacement between the classical trajectories for the center of the wave packets in Fig.~\ref{fig:interferometer1}:
\begin{equation}
\delta \boldsymbol{\chi} (t) =
(\mathcal{T}_\text{ret} \cdot \delta\boldsymbol{\mathcal{F}}_\text{lp}) (t)
\label{eq:displacement_solution},
\end{equation}
where $\delta\boldsymbol{\mathcal{F}}_\text{lp}$ simply corresponds to the difference between the laser kicks, as given by Eq.~\eqref{eq:laser_force}, for the two trajectories.

Making use of Eq.~\eqref{eq:transition_linear}, one obtains from Eq.~\eqref{eq:displacement_solution} the following relative displacement at each exit port for a Mach-Zehnder interferometer (depicted in Fig.~\ref{fig:interferometer1}), where the effective momentum transfer $\hbar \mathbf{k}$ is the same for all the laser pulses and induces a recoil velocity $\mathbf{v}_\text{rec} = \hbar \mathbf{k} / m$:
\begin{align}
\delta \boldsymbol{\mathcal{R}} &\approx - \mathbf{v}_\text{rec} \, \delta T
+ \left(\Gamma\, T^2 \right) \mathbf{v}_\text{rec} \, T
\label{eq:displacement_mzi_r} , \\
\delta \boldsymbol{\mathcal{P}} & \approx
\left(\Gamma\, T^2 \right) m \mathbf{v}_\text{rec}
\label{eq:displacement_mzi_p} ,
\end{align}
with time separations between the central $\pi$ pulse and the initial and final $\pi/2$ pulses of $T$ and $T + \delta T$ respectively and where higher-order terms in $\delta T / T$ and $(\Gamma\, T^2)$ have been neglected.

\subsection{Phase shift}
\label{sec:phase_shift}

\begin{figure}[h]
\begin{center}
\includegraphics[width=8cm]{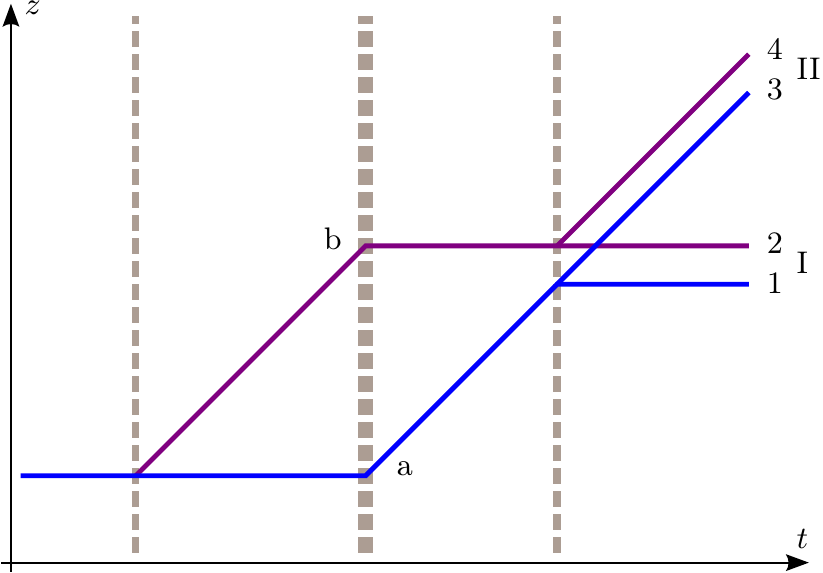}
\end{center}
\caption{Classical trajectories characterizing the motion of the wave packets in a Mach-Zehnder interferometer. The two branches of the interferometer are labeled as $a$ and $b$, whereas the trajectories of the two interfering wave packets at each exit port are labeled as $\{1,2\}$ and $\{3,4\}$ respectively.}
\label{fig:interferometer1}
\end{figure}

As explained above, the evolution of the corresponding wave packet along each branch is given by Eqs.~\eqref{eq:state_evol} and \eqref{eq:total_phase} with a displacement $\boldsymbol{\chi}(t)$ representing the evolution of its central position and momentum and following classical phase-space trajectories including the kicks from the laser pulses. This is indeed the full picture for mirror pulses ($\pi$ pulses). Beam-splitter pulses ($\pi/2$ pulses), however, create a superposition of two wave packets with different central momenta and, besides including the appropriate $1/\sqrt{2}$ normalization factor, one needs to consider also the evolution along each new branch from that point on.
This is illustrated for the example of a Mach-Zehnder interferometer (which consists of a $\pi/2$\,--\,$\pi$\,--\,$\pi/2$ pulse sequence) in Fig.~\ref{fig:interferometer1}, where the central trajectories for the various wave packets are depicted.

Assuming that the two exit ports are perfectly distinguishable due to spatial separation or internal-state labeling, one gets the following result for the state at the exit port I (an analogous result holds for exit port II):
\begin{align}
|\psi_\text{I} (t)\rangle &=
\frac{1}{2} \Big[ e^{i \Phi_1} \hat{\mathcal{D}}(\boldsymbol{\chi}_1)  |\psi_\text{c} (t) \rangle
+ e^{i \Phi_2} \hat{\mathcal{D}}(\boldsymbol{\chi}_2)  |\psi_\text{c} (t) \rangle \Big] \nonumber \\
&= \frac{1}{2} e^{i \Phi_1} \hat{\mathcal{D}}(\boldsymbol{\chi}_1)
\Big[ 1 + e^{i \delta\Phi} \hat{\mathcal{D}}(\delta\boldsymbol{\chi}) \Big] |\psi_\text{c} (t) \rangle
\label{4},
\end{align}
with $\delta\boldsymbol{\chi} = \boldsymbol{\chi}_2 - \boldsymbol{\chi}_1$ and 
$\delta\Phi = \Phi_2 - \Phi_1 + \boldsymbol{\chi}_1^\text{T} J \, \boldsymbol{\chi}_2 / 2 \hbar$, where the extra phase arises from the composition of displacement operators according to Eq.~\eqref{eq:composition}.
Hence, the probability of detection in port I is given by
\begin{equation}
\big\langle \psi_\text{I} (t) \big| \psi_\text{I} (t) \big\rangle
= \frac{1}{2} \big(1 + C \cos \delta\phi \big)
\label{eq:probability_I}.
\end{equation}
where
\begin{equation}
C = \Big| \big\langle \psi_\text{c} (t) \big| \hat{\mathcal{D}}(\delta\boldsymbol{\chi})
\big| \psi_\text{c} (t) \big\rangle \Big| \leq 1
\label{eq:contrast2},
\end{equation}
is known as the contrast and characterizes the amplitude of the oscillations in the detection probability as a function of the phase shift. Note that $\delta\phi$ denotes the phase shift $\delta\Phi$ plus the phase of $\big\langle \psi_\text{c} (t) \big| \hat{\mathcal{D}}(\delta\boldsymbol{\chi}) \big| \psi_\text{c} (t) \big\rangle$, which is in general a complex quantity.

Making use of Eq.~\eqref{eq:total_phase} together with the equations of motion derived from a general quadratic Hamiltonian and the properties of their associated transition matrix $ \mathcal{T}(t',t'')$,
a compact derivation of the general formula for the phase shift was provided in Ref.~\cite{roura14} and the following result was obtained:
\begin{align}
\delta\Phi(t) &= \delta\varphi - \frac{1}{\hbar} \int_{t_0}^t \! dt' \, \Bigg(
\Big[ \delta\boldsymbol{\mathcal{F}}_\text{lp}^\text{T}(t') + \delta\boldsymbol{\mathcal{G}}^\text{T}(t') \Big]
J\, \boldsymbol{\bar{\chi}} (t') \nonumber \\
& \qquad \qquad \qquad \qquad \;
+ \delta V_0 (t') \Bigg)
\label{eq:phase_shift2},
\end{align}
where $\boldsymbol{\bar{\chi}} (t')$ is given by Eq.~\eqref{eq:full_solution} with $\boldsymbol{\bar{\mathcal{F}}}_\text{lp}$ and $\boldsymbol{\bar{\mathcal{G}}}$ as sources; moreover, a bar over any quantity denotes the semisum of its values for the two branches, i.e.\ $\bar{\mathcal{A}} \equiv (\mathcal{A}_1 + \mathcal{A}_2)/2$, and similarly the difference is denoted by $\delta\mathcal{A} \equiv \mathcal{A}_2 - \mathcal{A}_1$ in all cases except for $\delta\Phi$, which is defined otherwise above.

We note that Eq.~\eqref{eq:phase_shift2} agrees with the general formula obtained by Antoine and Brod\'e in Ref.~\cite{antoine03b} and generalizes it to the case of possibly branch-dependent forces (corresponding to $\delta\boldsymbol{\mathcal{G}} \neq 0$).
Furthermore, it is sometimes convenient to write the result in an alternative way which explicitly displays all the dependence on the initial conditions. When doing so, Eq.~\eqref{eq:phase_shift2} becomes
\begin{align}
\delta\Phi(t) & = \delta\varphi - \frac{1}{\hbar} \int_{t_0}^t dt'\, \delta V_0 (t')
- \frac{1}{\hbar}\, \delta \boldsymbol{\chi}^\text{T}(t)\, J \, 
\mathcal{T}(t,t_0) \, \boldsymbol{\chi}_0
\nonumber \\
&\quad - \frac{1}{\hbar} \int_{t_0}^t dt' \int_{t_0}^{t'} dt''
\Big[ \delta\boldsymbol{\mathcal{F}}_\text{lp}^\text{T}(t') + \delta\boldsymbol{\mathcal{G}}^\text{T}(t') \Big] J\,
\nonumber\\
& \quad \quad \quad \quad \times \mathcal{T}(t',t'') \,
\Big[ \boldsymbol{\bar{\mathcal{F}}}_\text{lp}(t'') + \boldsymbol{\bar{\mathcal{G}}}(t'') \Big]
\label{eq:phase_shift3}.
\end{align}
In addition to the term containing all the dependence on the central position and momentum of the initial wave packet, which plays a central role in Sec.~\ref{sec:co-location}, the remaining contributions can be straightforwardly evaluated by computing the integrals in the last term of Eq.~\eqref{eq:phase_shift3} and making use of Eqs.~\eqref{eq:laser_force}-\eqref{eq:g_force} together with \eqref{eq:transition_linear}. The standard results are then recovered.
For example, assuming $\delta\boldsymbol{\mathcal{G}} = 0$ as well as time-independent $\mathbf{g}$ and $\Gamma$ for simplicity, to lowest order in $\Gamma$ the combination of $\delta\boldsymbol{\mathcal{F}}_\text{lp}$ and $\boldsymbol{\bar{\mathcal{G}}}$ gives the usual $\mathbf{k}_\text{eff}^\text{T}\, \mathbf{g}\, T^2$ term, whereas the combination of $\delta\boldsymbol{\mathcal{F}}_\text{lp}$ and $\boldsymbol{\bar{\mathcal{F}}}_\text{lp}$ gives 
$\mathbf{k}_\text{eff}^\text{T}\, (\Gamma\, T^2)\, \mathbf{v}_\text{rec} T$, which is often known as the ``photon-recoil term'' and cancels out when the $\mathbf{k}$-reversal method or double diffraction are employed.
The effects of rotations, in turn, can be easily included by working in a non-rotating frame and considering a rotating $\mathbf{k}_\text{eff}$ for the laser pulses, as thoroughly studied in Ref.~\cite{kleinert15}.

It should be stressed that although we have focused on the Hamiltonian~\eqref{eq:hamiltonian}, the results in this appendix can be directly applied to a general quadratic Hamiltonian (including also general linear terms). In particular, the results can be straightforwardly applied to calculations in rotating frames, where new quadratic terms associated with the Coriolis and centrifugal forces arise.

\appsection{Contrast, Wigner function and Gaussian states}
\label{sec:contrast_wigner_gaussian}

As explained in Ref.~\cite{roura14}, the results of Appendix~\ref{sec:state_evolution} can be naturally generalized to mixed states, which can model for instance thermal clouds or a stochastic distribution from shot to shot of the initial values for the central position and velocity of the atomic wave packet. The centered state is then given by a density matrix $\hat{\rho}_\text{c} (t)$ and Eq.~\eqref{eq:contrast} becomes
\begin{equation}
C = \Big|\, \mathrm{Tr} \Big[ \hat{\mathcal{D}}(\delta\boldsymbol{\chi})\, \hat{\rho}_\text{c} (t) \Big] \,\Big|
\label{eq:contrast_mixed} .
\end{equation}
Furthermore, it is particularly interesting to analyze the interferometer contrast in terms of the Wigner function~\cite{hillary84}, a normalized and real-valued phase space distribution which plays a central role in the phase-space formulation of quantum mechanics \cite{schleich} and is defined as
\begin{equation}
W(\mathbf{x},\mathbf{p}) = \int \frac{d^3\Delta}{(2\pi\hbar)^3} \,
e^{i \mathbf{p}^\mathrm{T} \boldsymbol{\Delta}/ \hbar} \,
\big\langle \mathbf{x} - \boldsymbol{\Delta}/2 \big| \, \hat{\rho}_\text{c} \,
\big| \mathbf{x} + \boldsymbol{\Delta}/2 \big\rangle
\label{eq:wigner_def}.
\end{equation}
The expression for the contrast takes then the following suggestive form:
\begin{align}
C &= \int d^3x' \int d^3p' \, W(\mathbf{x}',\mathbf{p}';t) \,
e^{-\frac{i}{\hbar} \delta \boldsymbol{\chi}^\text{T} J \, \boldsymbol{\xi}'}  \nonumber \\
&= \int d^3x \int d^3p \, W(\mathbf{x},\mathbf{p};t_0) \,
e^{-\frac{i}{\hbar} \delta \boldsymbol{\chi}_0^\text{T} J \, \boldsymbol{\xi}}
\label{eq:contrast_wigner2},
\end{align}
where $\delta \boldsymbol{\chi}_0 \equiv \mathcal{T}^{-1} (t,t_0) \, \delta \boldsymbol{\chi}$ with the transition matrix $\mathcal{T} (t_2,t_1)$ defined in Appendix~\ref{sec:classical_trajectories}.
In the second equality we introduced the change of variables $\boldsymbol{\xi}' = \mathcal{T} (t,t_0) \, \boldsymbol{\xi}$, took into account that for quadratic potentials the Wigner function evolves exactly in the same way as a classical phase-space distribution and used the relation $\mathcal{T}^\mathrm{T} (t,t_0) \, J \, \mathcal{T} (t,t_0)=J$ satisfied by transition matrices associated with equations of motion derived from quadratic Hamiltonians. 

For time-independent gravity gradients one can use the exact result for the transition matrix in Eq.~\eqref{eq:transition_exact}, but given Earth's gravity gradient, $\Gamma_{zz} \approx 3 \times 10^{-6}\, \text{s}^{-2}$, neglecting higher-order terms in $(\Gamma\, T^2)$ is in most situations of interest an excellent approximation and makes expressions simpler and more transparent. 
This means that 
one can neglect the effect of gravity gradients in the evolution of the Wigner function for the centered state when calculating the contrast through Eq.~\eqref{eq:contrast_wigner2} or, equivalently, use the transition matrix for a free particle [neglecting the terms involving $\Gamma$ in Eq.~\eqref{eq:transition_linear}] in the relation between $\delta \boldsymbol{\chi}_0$ and $\delta \boldsymbol{\chi}$ , so that it becomes:
\begin{align}
\delta \boldsymbol{\mathcal{R}}_0 &= \delta \boldsymbol{\mathcal{R}}
- \frac{\delta \boldsymbol{\mathcal{P}}}{m} \, (t-t_0)
\label{eq:deltaR_free}, \\
\delta \boldsymbol{\mathcal{P}}_0 &= \delta \boldsymbol{\mathcal{P}}
\label{eq:deltaP_free}.
\end{align}

\subsection{Gaussian states}
\label{sec:gaussian_states}

For a general Gaussian state (either pure or mixed) the Wigner function takes the form
\begin{equation}
W(\mathbf{x},\mathbf{p};t_0) = (2\pi)^{-3} (\det{\Sigma})^{-\frac{1}{2}} \,
e^{-\frac{1}{2} \boldsymbol{\xi}^\text{T} \Sigma^{-1} \boldsymbol{\xi}}
\label{eq:wigner_gaussian},
\end{equation}
where $\boldsymbol{\xi} = (\mathbf{x},\mathbf{p})^\text{T}$ and $\Sigma$ is the phase-space covariance matrix, which is directly related to the Weyl-ordered two-point functions:
\begin{equation}
\Sigma_{ij} = \frac{1}{2} \left\langle \hat{\xi}_i \hat{\xi}_j + \hat{\xi}_j \hat{\xi}_i \right\rangle
= \left( \begin{array}{cc} \Sigma_{xx} & \Sigma_{xp} \\
\Sigma_{xp}^\text{T} & \Sigma_{pp}
\end{array} \right)_{ij}
\label{eq:covariance}.
\end{equation}
$\Sigma_{xx}$, $\Sigma_{xp}$ and $\Sigma_{pp}$ are $3\times3$ matrices that can be regarded as blocks of the $6\times6$ covariance matrix $\Sigma$, which is symmetric and positive definite.

From Eqs.~\eqref{eq:contrast_wigner2} and \eqref{eq:wigner_gaussian} we get the following result for the contrast defined in Eq.~\eqref{eq:contrast_mixed}:
\begin{equation}
C = e^{-\frac{1}{2 \hbar^2} \delta \boldsymbol{\chi}_0^\text{T} J^\text{T} \Sigma \,
J \, \delta \boldsymbol{\chi}_0}
\label{eq:contrast_gaussian2}.
\end{equation}
Up to a factor $1/2\hbar^2$ the exponent can be rewritten as
\begin{equation}
- \delta \boldsymbol{\mathcal{R}}_0^\text{(s)\,T} \Sigma_{pp} \,
\delta \boldsymbol{\mathcal{R}}_0^\text{(s)}
- \delta \boldsymbol{\mathcal{P}}_0^\text{T}
\Big( \Sigma_{xx} - \Sigma_{xp}\, \Sigma_{pp}^{-1} \, \Sigma_{xp}^\text{T} \Big)
\delta \boldsymbol{\mathcal{P}}_0
\label{eq:contrast_exponent},
\end{equation}
with $\delta \boldsymbol{\mathcal{R}}_0^\text{(s)} = \delta \boldsymbol{\mathcal{R}}_0
- \big( \Sigma_{pp}^{-1}\, \Sigma_{xp}^\text{T} \big)\,  \delta \boldsymbol{\mathcal{P}}_0$, which will be convenient for our discussion below on the mitigation strategy.

\subsection{Mitigation strategy}
\label{sec:mitigation_strategy}

An effective mitigation strategy against the loss of contrast due to gravity gradients was proposed in Ref.~\cite{roura14}. We illustrate it here with the example of Gaussian states. Let us consider the case of aligned gravity gradients, where the direction of $\mathbf{v}_\text{rec}$ coincides with a principal axis of the tensor $\Gamma$ \comment{and is also aligned with the covariance matrix $\Sigma$}.
Making use of Eqs.~\eqref{eq:displacement_mzi_r}-\eqref{eq:displacement_mzi_p} combined with Eqs.~\eqref{eq:deltaR_free}-\eqref{eq:deltaP_free}, one can see that the quantity $\delta \boldsymbol{\mathcal{R}}_0^\text{(s)}$ introduced above vanishes if one changes the time at which the last pulse is applied by
\comment{
\begin{equation}
\delta T = -(T+T_0) \big( \Gamma_{\parallel}\, T^2 \big)
- \big( \Sigma_{p_\|p_\|}^{-1}\, \Sigma_{x_\|p_\|}^\text{T} \big) / |\mathbf{v}_\text{rec}|
\label{eq:timing} ,
\end{equation}
}%
where $\Gamma_{\parallel}$ is the eigenvalue of the $\Gamma$ tensor along the direction of $\mathbf{v}_\text{rec}$
\comment{(similarly, the last term involves the components of $\Sigma_{pp}^{-1}$ and $\Sigma_{xx}$ along that direction)}
and we have considered a total time $t-t_0 = 2T+T_0$, with $T$ being  half the interferometer time and $T_0$ the time from $t_0$ till the first beam splitter.
It is clear from expression~\eqref{eq:contrast_exponent} that this choice maximizes the exponent in Eq.~\eqref{eq:contrast_gaussian2} and, therefore, minimizes the loss of contrast.

\begin{figure}[h]
\begin{center}
\includegraphics[width=8.5cm]{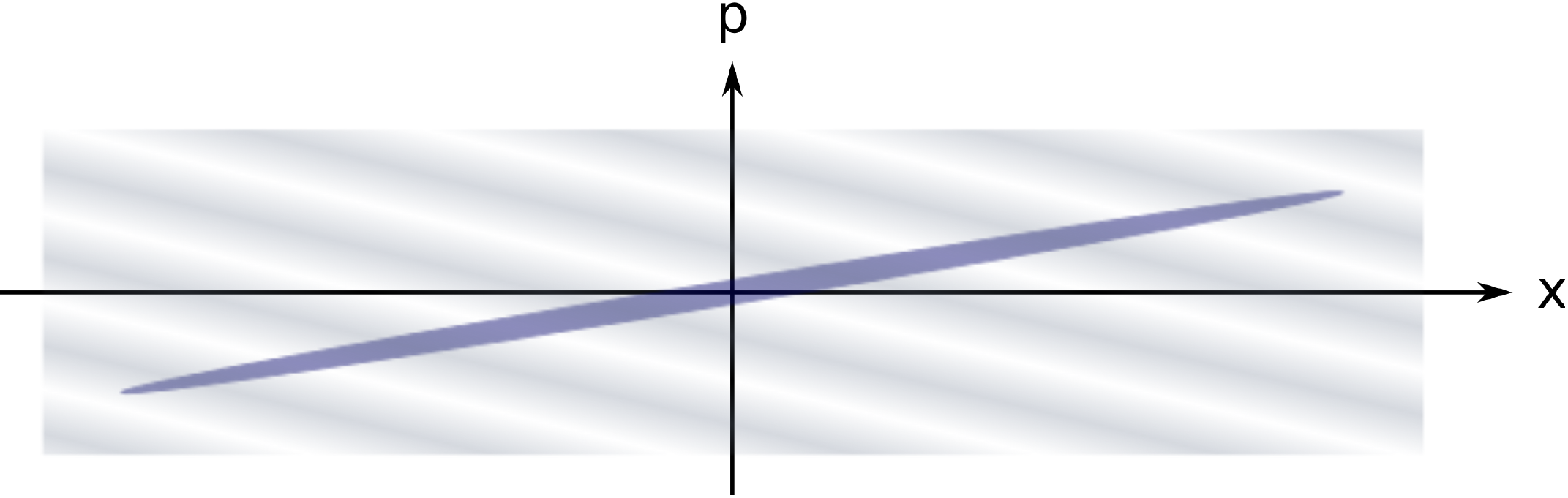}
\\ \vspace*{2.0em}
\includegraphics[width=8.5cm]{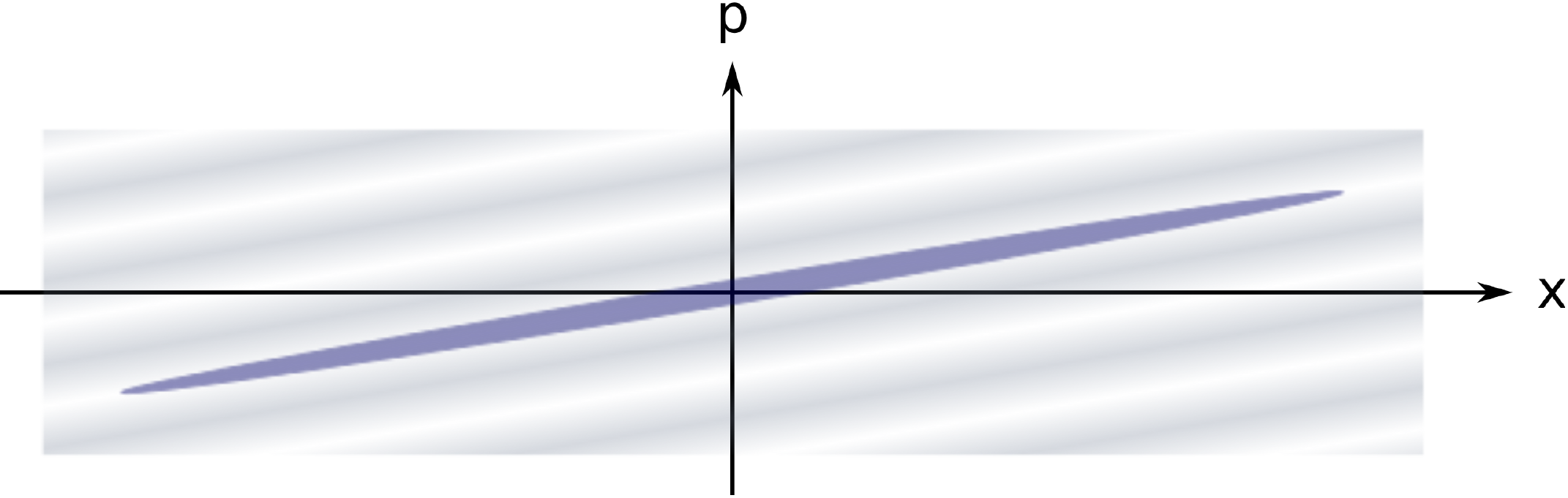}
\end{center}
\caption{Schematic representation of the main support of the Wigner function (blue ellipse) and of the phase factor in Eq.~\eqref{eq:contrast_wigner2} (grey color-gradient bands), both without mitigation strategy (a) and with the right choice of $\delta T$ so that the oscillations associated with the phase factor become aligned with the Wigner function (b).}
\label{fig:contrast_wigner}
\end{figure}

It is instructive to analyze this mitigation strategy in light of Eq.~\eqref{eq:contrast_wigner2}. The loss of contrast can be understood as a result of the wash-out effect due to the phase factor when evaluating the phase-space integral, and corresponds to the situation depicted in Fig.~\ref{fig:contrast_wigner}a. This effect can be reduced by decreasing both $\Sigma_{xx}$ and $\Sigma_{pp}$, but this is ultimately limited by Heisenberg's uncertainty principle, as discussed in Sec.~\ref{sec:challenges}. The change of $\delta \boldsymbol{\mathcal{R}}$ induced by $\delta T$ in Eq.~\eqref{eq:timing}, in contrast, corresponds to changing the orientation of the oscillations associated with the phase factor in Eq.~\eqref{eq:contrast_wigner2} so that they become aligned with the Wigner function and the wash-out effect is minimized, as shown in Fig.~\ref{fig:contrast_wigner}b.
In fact, this resembles the use of squeezed states to beat the standard quantum limit in optical interferometers. Instead of squeezing the state along the right direction in phase space, here the observable is modified to achieve an analogous effect.

In order to discuss how effective this mitigation strategy can be, we return to the example of Gaussian states and focus on the one-dimensional case for simplicity. (This still accounts for the three-dimensional case when $\mathbf{v}_\text{rec}$ and the principal axes of $\Gamma$ and $\Sigma$ are aligned.) After applying the mitigation strategy, we have $\delta \mathcal{R}_0^\text{(s)} = 0$ and are left with the following result for the contrast:
\begin{equation}
C = \exp \Big(\! -\frac{1}{2 \hbar^2} \frac{\det \Sigma}{\Sigma_{pp}} \, \delta \mathcal{P}_0^2
\Big)
\label{eq:contrast_gaussian3},
\end{equation}
where we have written in terms of $(\det \Sigma)$ the remaining contribution in expression~\eqref{eq:contrast_exponent}. 
Furthermore, it is useful to consider the following inequality satisfied by the exponent:
\begin{equation}
- \frac{1}{2 \hbar^2} \frac{\det \Sigma}{\Sigma_{pp}} \, \delta \mathcal{P}_0^2
\leq - \frac{1}{8} \frac{\delta \mathcal{P}_0^2}{\Sigma_{pp}}
\label{eq:dP^2} ,
\end{equation}
which follows from the inequality $(\det{\Sigma}) \geq \hbar^2 / 4$ for Gaussian states and where the equality holds for pure states and the strict inequality for mixed ones. 
For pure states and parameter ranges like those of the proposed STE-QUEST mission \cite{aguilera14} the exponent is much smaller than one in absolute value and there is hardly any loss of contrast, as shown in Ref.~\cite{roura14}, where the case of BECs in the Thomas-Fermi regime was also quantitatively investigated.
Qualitatively the need for having $\delta \mathcal{P}_0$ much smaller than the momentum spread can be understood as follows: for pure states the larger the momentum spread, the narrower the support becomes along the direction of the oscillations generated by the exponential factor in Eq.~\eqref{eq:contrast_wigner2}. This would eventually become a limitation to further enhancements of the sensitivity by 2 or 3 orders of magnitude compared to the proposed target for STE-QUEST
\comment{(and implying an increase of $\delta \mathcal{P}_0$ by the same amount)}
because the momentum spread needs to be kept sufficiently small to guarantee good diffraction efficiencies and to prevent the size of the atomic cloud from growing excessively. Such a limitation becomes more severe for mixed states, as seen from expression~\eqref{eq:dP^2} and implied by their larger phase-space volume%
\footnote{Alternatively, by regarding a mixed state as an ensemble of pure states with different central positions and momenta, the higher loss of contrast for mixed states can be understood as a dephasing effect between different members of the ensemble due to the dependence of the phase shift on the initial value of the central position and momentum \cite{roura14}.}.
In particular this would affect the use of ultracold atoms close to quantum degeneracy but with a negligible condensate fraction.
In contrast, the novel scheme presented in Sec.~\ref{sec:co-location} does not suffer from these shortcomings
\comment{(nor the need for $\Sigma$ being aligned with $\mathbf{v}_\text{rec}$)}
in addition to overcoming the initial co-location problem.

We conclude by mentioning that the strategy outlined above can be extended to the case of non-aligned gravity gradients (and of non-aligned $\Sigma$ too) by combining the adjustment of the pulse timing with the use of the same tip-tilt mirror which is employed to compensate the effect of rotations.
Further details can be found in  Ref.~\cite{roura14}.




\end{document}

%% file: figure2_pdf.tex
\begingroup%
  \makeatletter%
  \providecommand\color[2][]{%
    \errmessage{(Inkscape) Color is used for the text in Inkscape, but the package 'color.sty' is not loaded}%
    \renewcommand\color[2][]{}%
  }%
  \providecommand\transparent[1]{%
    \errmessage{(Inkscape) Transparency is used (non-zero) for the text in Inkscape, but the package 'transparent.sty' is not loaded}%
    \renewcommand\transparent[1]{}%
  }%
  \providecommand\rotatebox[2]{#2}%
  \ifx\svgwidth\undefined%
    \setlength{\unitlength}{1329.6015625bp}%
    \ifx\svgscale\undefined%
      \relax%
    \else%
      \setlength{\unitlength}{\unitlength * \real{\svgscale}}%
    \fi%
  \else%
    \setlength{\unitlength}{\svgwidth}%
  \fi%
  \global\let\svgwidth\undefined%
  \global\let\svgscale\undefined%
  \makeatother%
  \begin{picture}(1,0.70055724)%
    \put(0,0){\includegraphics[width=\unitlength]{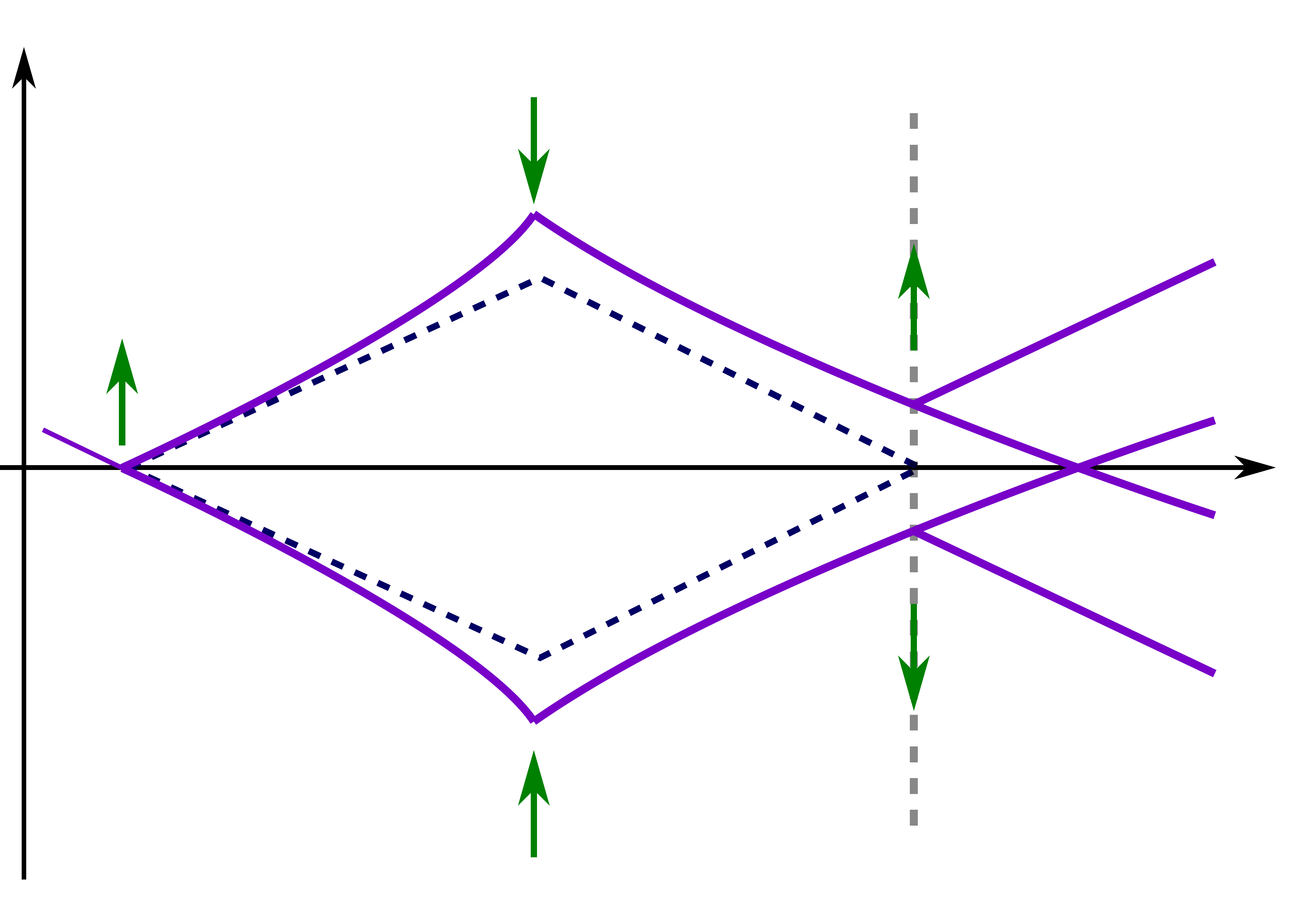}}%
    \put(0.00977737,0.71409279){\color[rgb]{0,0,0}\makebox(0,0)[lt]{\begin{minipage}{0.05214596\unitlength}\raggedright $z$\end{minipage}}}%
    \put(0.9889178,0.36631943){\color[rgb]{0,0,0}\makebox(0,0)[lt]{\begin{minipage}{0.08423576\unitlength}\raggedright $t$\end{minipage}}}%
    \put(0.07475924,0.50952022){\color[rgb]{0,0.5,0}\makebox(0,0)[lt]{\begin{minipage}{0.04612912\unitlength}\raggedright $\mathbf{k}_\text{eff}$ \end{minipage}}}%
    \put(0.38883829,0.69002542){\color[rgb]{0,0.5,0}\makebox(0,0)[lt]{\begin{minipage}{0.04612912\unitlength}\raggedright $\mathbf{k}_\text{eff}$ \end{minipage}}}%
    \put(0.38883829,0.03298649){\color[rgb]{0,0.5,0}\makebox(0,0)[lt]{\begin{minipage}{0.04612912\unitlength}\raggedright $\mathbf{k}_\text{eff}$ \end{minipage}}}%
    \put(0.67523988,0.67558501){\color[rgb]{0,0.5,0}\makebox(0,0)[lt]{\begin{minipage}{0.04612912\unitlength}\raggedright $\mathbf{k}_\text{eff}$ \end{minipage}}}%
    \put(0.67764662,0.05585048){\color[rgb]{0,0.5,0}\makebox(0,0)[lt]{\begin{minipage}{0.04612912\unitlength}\raggedright $\mathbf{k}_\text{eff}$ \end{minipage}}}%
  \end{picture}%
\endgroup%

%% file: figure3_pdf.tex
\begingroup%
  \makeatletter%
  \providecommand\color[2][]{%
    \errmessage{(Inkscape) Color is used for the text in Inkscape, but the package 'color.sty' is not loaded}%
    \renewcommand\color[2][]{}%
  }%
  \providecommand\transparent[1]{%
    \errmessage{(Inkscape) Transparency is used (non-zero) for the text in Inkscape, but the package 'transparent.sty' is not loaded}%
    \renewcommand\transparent[1]{}%
  }%
  \providecommand\rotatebox[2]{#2}%
  \ifx\svgwidth\undefined%
    \setlength{\unitlength}{1327.2015625bp}%
    \ifx\svgscale\undefined%
      \relax%
    \else%
      \setlength{\unitlength}{\unitlength * \real{\svgscale}}%
    \fi%
  \else%
    \setlength{\unitlength}{\svgwidth}%
  \fi%
  \global\let\svgwidth\undefined%
  \global\let\svgscale\undefined%
  \makeatother%
  \begin{picture}(1,0.69281112)%
    \put(0,0){\includegraphics[width=\unitlength]{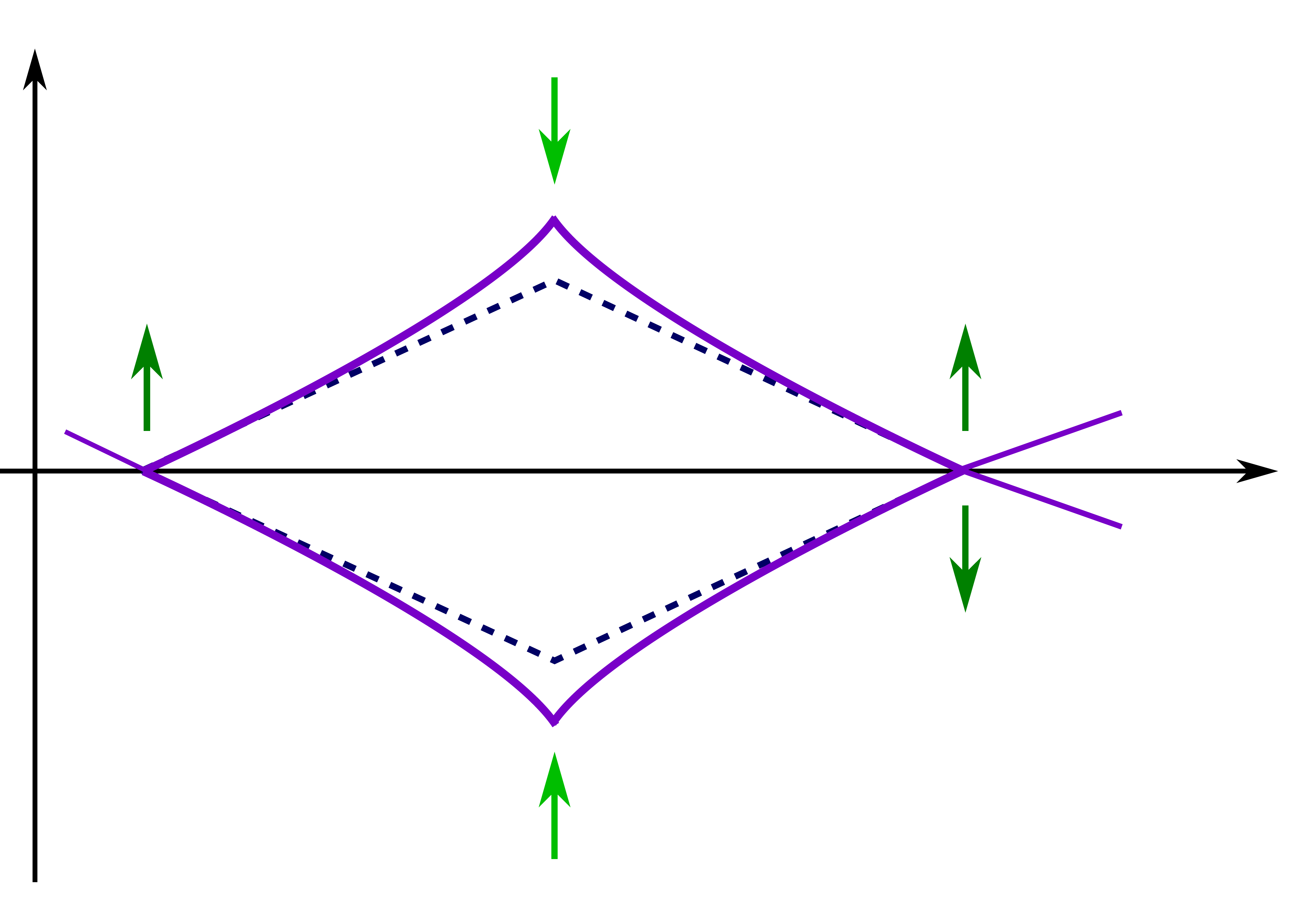}}%
    \put(0.01801749,0.70637404){\color[rgb]{0,0,0}\makebox(0,0)[lt]{\begin{minipage}{0.02782026\unitlength}\raggedright $z$\end{minipage}}}%
    \put(0.98891321,0.35537525){\color[rgb]{0,0,0}\makebox(0,0)[lt]{\begin{minipage}{0.0695506\unitlength}\raggedright $t$\end{minipage}}}%
    \put(0.09560504,0.50127699){\color[rgb]{0,0.5,0}\makebox(0,0)[lt]{\begin{minipage}{0.04173038\unitlength}\raggedright $\mathbf{k}_\text{eff}$\end{minipage}}}%
    \put(0.71695484,0.50093696){\color[rgb]{0,0.5,0}\makebox(0,0)[lt]{\begin{minipage}{0.05718607\unitlength}\raggedright $\mathbf{k}_\text{eff}$\end{minipage}}}%
    \put(0.342,0.68825996){\color[rgb]{0,0.745,0}\makebox(0,0)[lt]{\begin{minipage}{0.19\unitlength}\raggedright $\mathbf{k}_\text{eff}%
    + \Delta\mathbf{k}_\text{eff}$\end{minipage}}}%
    \put(0.71695484,0.21160638){\color[rgb]{0,0.5,0}\makebox(0,0)[lt]{\begin{minipage}{0.05718607\unitlength}\raggedright $\mathbf{k}_\text{eff}$\end{minipage}}}%
    \put(0.342,0.02){\color[rgb]{0,0.745,0}\makebox(0,0)[lt]{\begin{minipage}{0.19\unitlength}\raggedright $\mathbf{k}_\text{eff}%
    + \Delta\mathbf{k}_\text{eff}$\end{minipage}}}%
  \end{picture}%
\endgroup%

%% file: preprint3.bbl
\begin{thebibliography}{39}%
\makeatletter
\providecommand \@ifxundefined [1]{%
 \@ifx{#1\undefined}
}%
\providecommand \@ifnum [1]{%
 \ifnum #1\expandafter \@firstoftwo
 \else \expandafter \@secondoftwo
 \fi
}%
\providecommand \@ifx [1]{%
 \ifx #1\expandafter \@firstoftwo
 \else \expandafter \@secondoftwo
 \fi
}%
\providecommand \natexlab [1]{#1}%
\providecommand \enquote  [1]{``#1''}%
\providecommand \bibnamefont  [1]{#1}%
\providecommand \bibfnamefont [1]{#1}%
\providecommand \citenamefont [1]{#1}%
\providecommand \href@noop [0]{\@secondoftwo}%
\providecommand \href [0]{\begingroup \@sanitize@url \@href}%
\providecommand \@href[1]{\@@startlink{#1}\@@href}%
\providecommand \@@href[1]{\endgroup#1\@@endlink}%
\providecommand \@sanitize@url [0]{\catcode `\\12\catcode `\$12\catcode
  `\&12\catcode `\#12\catcode `\^12\catcode `\_12\catcode `\%12\relax}%
\providecommand \@@startlink[1]{}%
\providecommand \@@endlink[0]{}%
\providecommand \url  [0]{\begingroup\@sanitize@url \@url }%
\providecommand \@url [1]{\endgroup\@href {#1}{\urlprefix }}%
\providecommand \urlprefix  [0]{URL }%
\providecommand \Eprint [0]{\href }%
\providecommand \doibase [0]{http://dx.doi.org/}%
\providecommand \selectlanguage [0]{\@gobble}%
\providecommand \bibinfo  [0]{\@secondoftwo}%
\providecommand \bibfield  [0]{\@secondoftwo}%
\providecommand \translation [1]{[#1]}%
\providecommand \BibitemOpen [0]{}%
\providecommand \bibitemStop [0]{}%
\providecommand \bibitemNoStop [0]{.\EOS\space}%
\providecommand \EOS [0]{\spacefactor3000\relax}%
\providecommand \BibitemShut  [1]{\csname bibitem#1\endcsname}%
\let\auto@bib@innerbib\@empty
\bibitem [{\citenamefont {Will}(2014)}]{will14}%
  \BibitemOpen
  \bibfield  {author} {\bibinfo {author} {\bibfnamefont {C.~M.}\ \bibnamefont
  {Will}},\ }\href@noop {} {\bibfield  {journal} {\bibinfo  {journal} {Living
  Rev. Relativity}\ }\textbf {\bibinfo {volume} {17}} (\bibinfo {year}
  {2014})}\BibitemShut {NoStop}%
\bibitem [{\citenamefont {Kosteleck\'y}\ and\ \citenamefont
  {Tasson}(2011)}]{kostelecky11a}%
  \BibitemOpen
  \bibfield  {author} {\bibinfo {author} {\bibfnamefont {V.~A.}\ \bibnamefont
  {Kosteleck\'y}}\ and\ \bibinfo {author} {\bibfnamefont {J.~D.}\ \bibnamefont
  {Tasson}},\ }\href {\doibase 10.1103/PhysRevD.83.016013} {\bibfield
  {journal} {\bibinfo  {journal} {Phys. Rev. D}\ }\textbf {\bibinfo {volume}
  {83}},\ \bibinfo {pages} {016013} (\bibinfo {year} {2011})}\BibitemShut
  {NoStop}%
\bibitem [{\citenamefont {Damour}\ and\ \citenamefont
  {Donoghue}(2010)}]{damour10}%
  \BibitemOpen
  \bibfield  {author} {\bibinfo {author} {\bibfnamefont {T.}~\bibnamefont
  {Damour}}\ and\ \bibinfo {author} {\bibfnamefont {J.~F.}\ \bibnamefont
  {Donoghue}},\ }\href@noop {} {\bibfield  {journal} {\bibinfo  {journal}
  {Phys. Rev. D}\ }\textbf {\bibinfo {volume} {82}},\ \bibinfo {pages} {084033}
  (\bibinfo {year} {2010})}\BibitemShut {NoStop}%
\bibitem [{\citenamefont {Schlamminger}\ \emph {et~al.}(2008)\citenamefont
  {Schlamminger}, \citenamefont {Choi}, \citenamefont {Wagner}, \citenamefont
  {Gundlach},\ and\ \citenamefont {Adelberger}}]{schlamminger08}%
  \BibitemOpen
  \bibfield  {author} {\bibinfo {author} {\bibfnamefont {S.}~\bibnamefont
  {Schlamminger}}, \bibinfo {author} {\bibfnamefont {K.-Y.}\ \bibnamefont
  {Choi}}, \bibinfo {author} {\bibfnamefont {T.~A.}\ \bibnamefont {Wagner}},
  \bibinfo {author} {\bibfnamefont {J.~H.}\ \bibnamefont {Gundlach}}, \ and\
  \bibinfo {author} {\bibfnamefont {E.~G.}\ \bibnamefont {Adelberger}},\ }\href
  {\doibase 10.1103/PhysRevLett.100.041101} {\bibfield  {journal} {\bibinfo
  {journal} {Phys. Rev. Lett.}\ }\textbf {\bibinfo {volume} {100}},\ \bibinfo
  {pages} {041101} (\bibinfo {year} {2008})}\BibitemShut {NoStop}%
\bibitem [{\citenamefont {Touboul}\ \emph {et~al.}(2012)\citenamefont
  {Touboul}, \citenamefont {Métris}, \citenamefont {Lebat},\ and\
  \citenamefont {Robert}}]{touboul12}%
  \BibitemOpen
  \bibfield  {author} {\bibinfo {author} {\bibfnamefont {P.}~\bibnamefont
  {Touboul}}, \bibinfo {author} {\bibfnamefont {G.}~\bibnamefont {Métris}},
  \bibinfo {author} {\bibfnamefont {V.}~\bibnamefont {Lebat}}, \ and\ \bibinfo
  {author} {\bibfnamefont {A.}~\bibnamefont {Robert}},\ }\href
  {http://stacks.iop.org/0264-9381/29/i=18/a=184010} {\bibfield  {journal}
  {\bibinfo  {journal} {Classical and Quantum Gravity}\ }\textbf {\bibinfo
  {volume} {29}},\ \bibinfo {pages} {184010} (\bibinfo {year}
  {2012})}\BibitemShut {NoStop}%
\bibitem [{\citenamefont {Fray}\ \emph {et~al.}(2004)\citenamefont {Fray},
  \citenamefont {Diez}, \citenamefont {H\"{a}nsch},\ and\ \citenamefont
  {Weitz}}]{fray04}%
  \BibitemOpen
  \bibfield  {author} {\bibinfo {author} {\bibfnamefont {S.}~\bibnamefont
  {Fray}}, \bibinfo {author} {\bibfnamefont {C.~A.}\ \bibnamefont {Diez}},
  \bibinfo {author} {\bibfnamefont {T.~W.}\ \bibnamefont {H\"{a}nsch}}, \ and\
  \bibinfo {author} {\bibfnamefont {M.}~\bibnamefont {Weitz}},\ }\href
  {\doibase 10.1103/PhysRevLett.93.240404} {\bibfield  {journal} {\bibinfo
  {journal} {Phys. Rev. Lett.}\ }\textbf {\bibinfo {volume} {93}},\ \bibinfo
  {pages} {240404} (\bibinfo {year} {2004})}\BibitemShut {NoStop}%
\bibitem [{\citenamefont {Bonnin}\ \emph {et~al.}(2013)\citenamefont {Bonnin},
  \citenamefont {Zahzam}, \citenamefont {Bidel},\ and\ \citenamefont
  {Bresson}}]{bonnin13}%
  \BibitemOpen
  \bibfield  {author} {\bibinfo {author} {\bibfnamefont {A.}~\bibnamefont
  {Bonnin}}, \bibinfo {author} {\bibfnamefont {N.}~\bibnamefont {Zahzam}},
  \bibinfo {author} {\bibfnamefont {Y.}~\bibnamefont {Bidel}}, \ and\ \bibinfo
  {author} {\bibfnamefont {A.}~\bibnamefont {Bresson}},\ }\href {\doibase
  10.1103/PhysRevA.88.043615} {\bibfield  {journal} {\bibinfo  {journal} {Phys.
  Rev. A}\ }\textbf {\bibinfo {volume} {88}},\ \bibinfo {pages} {043615}
  (\bibinfo {year} {2013})}\BibitemShut {NoStop}%
\bibitem [{\citenamefont {Schlippert}\ \emph {et~al.}(2014)\citenamefont
  {Schlippert}, \citenamefont {Hartwig}, \citenamefont {Albers}, \citenamefont
  {Richardson}, \citenamefont {Schubert}, \citenamefont {Roura}, \citenamefont
  {Schleich}, \citenamefont {Ertmer},\ and\ \citenamefont
  {Rasel}}]{schlippert14}%
  \BibitemOpen
  \bibfield  {author} {\bibinfo {author} {\bibfnamefont {D.}~\bibnamefont
  {Schlippert}}, \bibinfo {author} {\bibfnamefont {J.}~\bibnamefont {Hartwig}},
  \bibinfo {author} {\bibfnamefont {H.}~\bibnamefont {Albers}}, \bibinfo
  {author} {\bibfnamefont {L.~L.}\ \bibnamefont {Richardson}}, \bibinfo
  {author} {\bibfnamefont {C.}~\bibnamefont {Schubert}}, \bibinfo {author}
  {\bibfnamefont {A.}~\bibnamefont {Roura}}, \bibinfo {author} {\bibfnamefont
  {W.~P.}\ \bibnamefont {Schleich}}, \bibinfo {author} {\bibfnamefont
  {W.}~\bibnamefont {Ertmer}}, \ and\ \bibinfo {author} {\bibfnamefont {E.~M.}\
  \bibnamefont {Rasel}},\ }\href {\doibase 10.1103/PhysRevLett.112.203002}
  {\bibfield  {journal} {\bibinfo  {journal} {Phys. Rev. Lett.}\ }\textbf
  {\bibinfo {volume} {112}},\ \bibinfo {pages} {203002} (\bibinfo {year}
  {2014})}\BibitemShut {NoStop}%
\bibitem [{\citenamefont {Zhou}\ \emph {et~al.}(2015)\citenamefont {Zhou},
  \citenamefont {Long}, \citenamefont {Tang}, \citenamefont {Chen},
  \citenamefont {Gao}, \citenamefont {Peng}, \citenamefont {Duan},
  \citenamefont {Zhong}, \citenamefont {Xiong}, \citenamefont {Wang},
  \citenamefont {Zhang},\ and\ \citenamefont {Zhan}}]{zhou15}%
  \BibitemOpen
  \bibfield  {author} {\bibinfo {author} {\bibfnamefont {L.}~\bibnamefont
  {Zhou}}, \bibinfo {author} {\bibfnamefont {S.}~\bibnamefont {Long}}, \bibinfo
  {author} {\bibfnamefont {B.}~\bibnamefont {Tang}}, \bibinfo {author}
  {\bibfnamefont {X.}~\bibnamefont {Chen}}, \bibinfo {author} {\bibfnamefont
  {F.}~\bibnamefont {Gao}}, \bibinfo {author} {\bibfnamefont {W.}~\bibnamefont
  {Peng}}, \bibinfo {author} {\bibfnamefont {W.}~\bibnamefont {Duan}}, \bibinfo
  {author} {\bibfnamefont {J.}~\bibnamefont {Zhong}}, \bibinfo {author}
  {\bibfnamefont {Z.}~\bibnamefont {Xiong}}, \bibinfo {author} {\bibfnamefont
  {J.}~\bibnamefont {Wang}}, \bibinfo {author} {\bibfnamefont {Y.}~\bibnamefont
  {Zhang}}, \ and\ \bibinfo {author} {\bibfnamefont {M.}~\bibnamefont {Zhan}},\
  }\href {\doibase 10.1103/PhysRevLett.115.013004} {\bibfield  {journal}
  {\bibinfo  {journal} {Phys. Rev. Lett.}\ }\textbf {\bibinfo {volume} {115}},\
  \bibinfo {pages} {013004} (\bibinfo {year} {2015})}\BibitemShut {NoStop}%
\bibitem [{\citenamefont {Hohensee}\ \emph {et~al.}(2013)\citenamefont
  {Hohensee}, \citenamefont {M\"uller},\ and\ \citenamefont
  {Wiringa}}]{hohensee13b}%
  \BibitemOpen
  \bibfield  {author} {\bibinfo {author} {\bibfnamefont {M.~A.}\ \bibnamefont
  {Hohensee}}, \bibinfo {author} {\bibfnamefont {H.}~\bibnamefont {M\"uller}},
  \ and\ \bibinfo {author} {\bibfnamefont {R.~B.}\ \bibnamefont {Wiringa}},\
  }\href {\doibase 10.1103/PhysRevLett.111.151102} {\bibfield  {journal}
  {\bibinfo  {journal} {Phys. Rev. Lett.}\ }\textbf {\bibinfo {volume} {111}},\
  \bibinfo {pages} {151102} (\bibinfo {year} {2013})}\BibitemShut {NoStop}%
\bibitem [{\citenamefont {Bord\'{e}}(1989)}]{borde89}%
  \BibitemOpen
  \bibfield  {author} {\bibinfo {author} {\bibfnamefont {C.~J.}\ \bibnamefont
  {Bord\'{e}}},\ }\href {\doibase 10.1016/0375-9601(89)90537-9} {\bibfield
  {journal} {\bibinfo  {journal} {Phys. Lett. A}\ }\textbf {\bibinfo {volume}
  {140}},\ \bibinfo {pages} {10 } (\bibinfo {year} {1989})}\BibitemShut
  {NoStop}%
\bibitem [{\citenamefont {Kasevich}\ and\ \citenamefont
  {Chu}(1991)}]{kasevich91}%
  \BibitemOpen
  \bibfield  {author} {\bibinfo {author} {\bibfnamefont {M.}~\bibnamefont
  {Kasevich}}\ and\ \bibinfo {author} {\bibfnamefont {S.}~\bibnamefont {Chu}},\
  }\href {\doibase 10.1103/PhysRevLett.67.181} {\bibfield  {journal} {\bibinfo
  {journal} {Phys. Rev. Lett.}\ }\textbf {\bibinfo {volume} {67}},\ \bibinfo
  {pages} {181} (\bibinfo {year} {1991})}\BibitemShut {NoStop}%
\bibitem [{\citenamefont {Peters}\ \emph {et~al.}(2001)\citenamefont {Peters},
  \citenamefont {Chung},\ and\ \citenamefont {Chu}}]{peters01}%
  \BibitemOpen
  \bibfield  {author} {\bibinfo {author} {\bibfnamefont {A.}~\bibnamefont
  {Peters}}, \bibinfo {author} {\bibfnamefont {K.~Y.}\ \bibnamefont {Chung}}, \
  and\ \bibinfo {author} {\bibfnamefont {S.}~\bibnamefont {Chu}},\ }\href
  {http://stacks.iop.org/0026-1394/38/i=1/a=4} {\bibfield  {journal} {\bibinfo
  {journal} {Metrologia}\ }\textbf {\bibinfo {volume} {38}},\ \bibinfo {pages}
  {25} (\bibinfo {year} {2001})}\BibitemShut {NoStop}%
\bibitem [{\citenamefont {Hu}\ \emph {et~al.}(2013)\citenamefont {Hu},
  \citenamefont {Sun}, \citenamefont {Duan}, \citenamefont {Zhou},
  \citenamefont {Chen}, \citenamefont {Zhan}, \citenamefont {Zhang},\ and\
  \citenamefont {Luo}}]{hu13}%
  \BibitemOpen
  \bibfield  {author} {\bibinfo {author} {\bibfnamefont {Z.-K.}\ \bibnamefont
  {Hu}}, \bibinfo {author} {\bibfnamefont {B.-L.}\ \bibnamefont {Sun}},
  \bibinfo {author} {\bibfnamefont {X.-C.}\ \bibnamefont {Duan}}, \bibinfo
  {author} {\bibfnamefont {M.-K.}\ \bibnamefont {Zhou}}, \bibinfo {author}
  {\bibfnamefont {L.-L.}\ \bibnamefont {Chen}}, \bibinfo {author}
  {\bibfnamefont {S.}~\bibnamefont {Zhan}}, \bibinfo {author} {\bibfnamefont
  {Q.-Z.}\ \bibnamefont {Zhang}}, \ and\ \bibinfo {author} {\bibfnamefont
  {J.}~\bibnamefont {Luo}},\ }\href@noop {} {\bibfield  {journal} {\bibinfo
  {journal} {Phys. Rev. A}\ }\textbf {\bibinfo {volume} {88}},\ \bibinfo
  {pages} {043610} (\bibinfo {year} {2013})}\BibitemShut {NoStop}%
\bibitem [{\citenamefont {Fixler}\ \emph {et~al.}(2007)\citenamefont {Fixler},
  \citenamefont {Foster}, \citenamefont {McGuirk},\ and\ \citenamefont
  {Kasevich}}]{fixler07}%
  \BibitemOpen
  \bibfield  {author} {\bibinfo {author} {\bibfnamefont {J.~B.}\ \bibnamefont
  {Fixler}}, \bibinfo {author} {\bibfnamefont {G.~T.}\ \bibnamefont {Foster}},
  \bibinfo {author} {\bibfnamefont {J.~M.}\ \bibnamefont {McGuirk}}, \ and\
  \bibinfo {author} {\bibfnamefont {M.~A.}\ \bibnamefont {Kasevich}},\ }\href
  {\doibase 10.1126/science.1135459} {\bibfield  {journal} {\bibinfo  {journal}
  {Science}\ }\textbf {\bibinfo {volume} {315}},\ \bibinfo {pages} {74}
  (\bibinfo {year} {2007})}\BibitemShut {NoStop}%
\bibitem [{\citenamefont {Hogan}\ \emph {et~al.}()\citenamefont {Hogan},
  \citenamefont {Johnson},\ and\ \citenamefont {Kasevich}}]{hogan08}%
  \BibitemOpen
  \bibfield  {author} {\bibinfo {author} {\bibfnamefont {J.~M.}\ \bibnamefont
  {Hogan}}, \bibinfo {author} {\bibfnamefont {D.~M.~S.}\ \bibnamefont
  {Johnson}}, \ and\ \bibinfo {author} {\bibfnamefont {M.~A.}\ \bibnamefont
  {Kasevich}},\ }\href {http://arxiv.org/abs/0806.3261} {\bibinfo  {journal}
  {\texttt{arXiv:0806.3261}}\ }\BibitemShut {NoStop}%
\bibitem [{\citenamefont {Varoquaux}\ \emph {et~al.}(2009)\citenamefont
  {Varoquaux}, \citenamefont {Nyman}, \citenamefont {Geiger}, \citenamefont
  {Cheinet}, \citenamefont {Landragin},\ and\ \citenamefont
  {Bouyer}}]{varoquaux09}%
  \BibitemOpen
\bibfield  {journal} {  }\bibfield  {author} {\bibinfo {author} {\bibfnamefont
  {G.}~\bibnamefont {Varoquaux}}, \bibinfo {author} {\bibfnamefont {R.~A.}\
  \bibnamefont {Nyman}}, \bibinfo {author} {\bibfnamefont {R.}~\bibnamefont
  {Geiger}}, \bibinfo {author} {\bibfnamefont {P.}~\bibnamefont {Cheinet}},
  \bibinfo {author} {\bibfnamefont {A.}~\bibnamefont {Landragin}}, \ and\
  \bibinfo {author} {\bibfnamefont {P.}~\bibnamefont {Bouyer}},\ }\href
  {http://stacks.iop.org/1367-2630/11/i=11/a=113010} {\bibfield  {journal}
  {\bibinfo  {journal} {New Journal of Physics}\ }\textbf {\bibinfo {volume}
  {11}},\ \bibinfo {pages} {113010} (\bibinfo {year} {2009})}\BibitemShut
  {NoStop}%
\bibitem [{\citenamefont {{Rosi}}\ \emph {et~al.}(2014)\citenamefont {{Rosi}},
  \citenamefont {{Sorrentino}}, \citenamefont {{Cacciapuoti}}, \citenamefont
  {{Prevedelli}},\ and\ \citenamefont {{Tino}}}]{rosi14}%
  \BibitemOpen
  \bibfield  {author} {\bibinfo {author} {\bibfnamefont {G.}~\bibnamefont
  {{Rosi}}}, \bibinfo {author} {\bibfnamefont {F.}~\bibnamefont
  {{Sorrentino}}}, \bibinfo {author} {\bibfnamefont {L.}~\bibnamefont
  {{Cacciapuoti}}}, \bibinfo {author} {\bibfnamefont {M.}~\bibnamefont
  {{Prevedelli}}}, \ and\ \bibinfo {author} {\bibfnamefont {G.~M.}\
  \bibnamefont {{Tino}}},\ }\href {\doibase 10.1038/nature13433} {\bibfield
  {journal} {\bibinfo  {journal} {Nature}\ }\textbf {\bibinfo {volume} {510}},\
  \bibinfo {pages} {518} (\bibinfo {year} {2014})}\BibitemShut {NoStop}%
\bibitem [{\citenamefont {Aguilera}\ \emph {et~al.}(2014)\citenamefont
  {Aguilera} \emph {et~al.}}]{aguilera14}%
  \BibitemOpen
  \bibfield  {author} {\bibinfo {author} {\bibfnamefont {D.~N.}\ \bibnamefont
  {Aguilera}} \emph {et~al.},\ }\href
  {http://stacks.iop.org/0264-9381/31/i=11/a=115010} {\bibfield  {journal}
  {\bibinfo  {journal} {Classical and Quantum Gravity}\ }\textbf {\bibinfo
  {volume} {31}},\ \bibinfo {pages} {115010} (\bibinfo {year}
  {2014})}\BibitemShut {NoStop}%
\bibitem [{\citenamefont {Roura}\ \emph {et~al.}(2014)\citenamefont {Roura},
  \citenamefont {Zeller},\ and\ \citenamefont {Schleich}}]{roura14}%
  \BibitemOpen
  \bibfield  {author} {\bibinfo {author} {\bibfnamefont {A.}~\bibnamefont
  {Roura}}, \bibinfo {author} {\bibfnamefont {W.}~\bibnamefont {Zeller}}, \
  and\ \bibinfo {author} {\bibfnamefont {W.~P.}\ \bibnamefont {Schleich}},\
  }\href {http://stacks.iop.org/1367-2630/16/i=12/a=123012} {\bibfield
  {journal} {\bibinfo  {journal} {New Journal of Physics}\ }\textbf {\bibinfo
  {volume} {16}},\ \bibinfo {pages} {123012} (\bibinfo {year}
  {2014})}\BibitemShut {NoStop}%
\bibitem [{\citenamefont {Nobili}()}]{nobili15}%
  \BibitemOpen
  \bibfield  {author} {\bibinfo {author} {\bibfnamefont {A.~M.}\ \bibnamefont
  {Nobili}},\ }\href@noop {} {\bibinfo  {journal} {\texttt{arXiv:1503.01074}}\
  }\BibitemShut {NoStop}%
\bibitem [{\citenamefont {Hillery}\ \emph {et~al.}(1984)\citenamefont
  {Hillery}, \citenamefont {O'Connell}, \citenamefont {Scully},\ and\
  \citenamefont {Wigner}}]{hillary84}%
  \BibitemOpen
\bibfield  {journal} {  }\bibfield  {author} {\bibinfo {author} {\bibfnamefont
  {M.}~\bibnamefont {Hillery}}, \bibinfo {author} {\bibfnamefont
  {R.}~\bibnamefont {O'Connell}}, \bibinfo {author} {\bibfnamefont
  {M.}~\bibnamefont {Scully}}, \ and\ \bibinfo {author} {\bibfnamefont
  {E.}~\bibnamefont {Wigner}},\ }\href {\doibase
  http://dx.doi.org/10.1016/0370-1573(84)90160-1} {\bibfield  {journal}
  {\bibinfo  {journal} {Physics Reports}\ }\textbf {\bibinfo {volume} {106}},\
  \bibinfo {pages} {121 } (\bibinfo {year} {1984})}\BibitemShut {NoStop}%
\bibitem [{\citenamefont {Dimopoulos}\ \emph {et~al.}(2007)\citenamefont
  {Dimopoulos}, \citenamefont {Graham}, \citenamefont {Hogan},\ and\
  \citenamefont {Kasevich}}]{dimopoulos07}%
  \BibitemOpen
  \bibfield  {author} {\bibinfo {author} {\bibfnamefont {S.}~\bibnamefont
  {Dimopoulos}}, \bibinfo {author} {\bibfnamefont {P.~W.}\ \bibnamefont
  {Graham}}, \bibinfo {author} {\bibfnamefont {J.~M.}\ \bibnamefont {Hogan}}, \
  and\ \bibinfo {author} {\bibfnamefont {M.~A.}\ \bibnamefont {Kasevich}},\
  }\href {\doibase 10.1103/PhysRevLett.98.111102} {\bibfield  {journal}
  {\bibinfo  {journal} {Phys. Rev. Lett.}\ }\textbf {\bibinfo {volume} {98}},\
  \bibinfo {pages} {111102} (\bibinfo {year} {2007})}\BibitemShut {NoStop}%
\bibitem [{\citenamefont {M\"{u}ntinga}\ \emph {et~al.}(2013)\citenamefont
  {M\"{u}ntinga}, \citenamefont {Ahlers}, \citenamefont {Krutzik},
  \citenamefont {Wenzlawski}, \citenamefont {Arnold}, \citenamefont {Becker},
  \citenamefont {Bongs}, \citenamefont {Dittus}, \citenamefont {Duncker},
  \citenamefont {Gaaloul}, \citenamefont {Gherasim}, \citenamefont {Giese},
  \citenamefont {Grzeschik}, \citenamefont {H\"{a}nsch}, \citenamefont
  {Hellmig}, \citenamefont {Herr}, \citenamefont {Herrmann}, \citenamefont
  {Kajari}, \citenamefont {Kleinert}, \citenamefont {L\"{a}mmerzahl},
  \citenamefont {Lewoczko-Adamczyk}, \citenamefont {Malcolm}, \citenamefont
  {Meyer}, \citenamefont {Nolte}, \citenamefont {Peters}, \citenamefont {Popp},
  \citenamefont {Reichel}, \citenamefont {Roura}, \citenamefont {Rudolph},
  \citenamefont {Schiemangk}, \citenamefont {Schneider}, \citenamefont
  {Seidel}, \citenamefont {Sengstock}, \citenamefont {Tamma}, \citenamefont
  {Valenzuela}, \citenamefont {Vogel}, \citenamefont {Walser}, \citenamefont
  {Wendrich}, \citenamefont {Windpassinger}, \citenamefont {Zeller},
  \citenamefont {van Zoest}, \citenamefont {Ertmer}, \citenamefont {Schleich},\
  and\ \citenamefont {Rasel}}]{muentinga13}%
  \BibitemOpen
  \bibfield  {author} {\bibinfo {author} {\bibfnamefont {H.}~\bibnamefont
  {M\"{u}ntinga}}, \bibinfo {author} {\bibfnamefont {H.}~\bibnamefont
  {Ahlers}}, \bibinfo {author} {\bibfnamefont {M.}~\bibnamefont {Krutzik}},
  \bibinfo {author} {\bibfnamefont {A.}~\bibnamefont {Wenzlawski}}, \bibinfo
  {author} {\bibfnamefont {S.}~\bibnamefont {Arnold}}, \bibinfo {author}
  {\bibfnamefont {D.}~\bibnamefont {Becker}}, \bibinfo {author} {\bibfnamefont
  {K.}~\bibnamefont {Bongs}}, \bibinfo {author} {\bibfnamefont
  {H.}~\bibnamefont {Dittus}}, \bibinfo {author} {\bibfnamefont
  {H.}~\bibnamefont {Duncker}}, \bibinfo {author} {\bibfnamefont
  {N.}~\bibnamefont {Gaaloul}}, \bibinfo {author} {\bibfnamefont
  {C.}~\bibnamefont {Gherasim}}, \bibinfo {author} {\bibfnamefont
  {E.}~\bibnamefont {Giese}}, \bibinfo {author} {\bibfnamefont
  {C.}~\bibnamefont {Grzeschik}}, \bibinfo {author} {\bibfnamefont {T.~W.}\
  \bibnamefont {H\"{a}nsch}}, \bibinfo {author} {\bibfnamefont
  {O.}~\bibnamefont {Hellmig}}, \bibinfo {author} {\bibfnamefont
  {W.}~\bibnamefont {Herr}}, \bibinfo {author} {\bibfnamefont {S.}~\bibnamefont
  {Herrmann}}, \bibinfo {author} {\bibfnamefont {E.}~\bibnamefont {Kajari}},
  \bibinfo {author} {\bibfnamefont {S.}~\bibnamefont {Kleinert}}, \bibinfo
  {author} {\bibfnamefont {C.}~\bibnamefont {L\"{a}mmerzahl}}, \bibinfo
  {author} {\bibfnamefont {W.}~\bibnamefont {Lewoczko-Adamczyk}}, \bibinfo
  {author} {\bibfnamefont {J.}~\bibnamefont {Malcolm}}, \bibinfo {author}
  {\bibfnamefont {N.}~\bibnamefont {Meyer}}, \bibinfo {author} {\bibfnamefont
  {R.}~\bibnamefont {Nolte}}, \bibinfo {author} {\bibfnamefont
  {A.}~\bibnamefont {Peters}}, \bibinfo {author} {\bibfnamefont
  {M.}~\bibnamefont {Popp}}, \bibinfo {author} {\bibfnamefont {J.}~\bibnamefont
  {Reichel}}, \bibinfo {author} {\bibfnamefont {A.}~\bibnamefont {Roura}},
  \bibinfo {author} {\bibfnamefont {J.}~\bibnamefont {Rudolph}}, \bibinfo
  {author} {\bibfnamefont {M.}~\bibnamefont {Schiemangk}}, \bibinfo {author}
  {\bibfnamefont {M.}~\bibnamefont {Schneider}}, \bibinfo {author}
  {\bibfnamefont {S.~T.}\ \bibnamefont {Seidel}}, \bibinfo {author}
  {\bibfnamefont {K.}~\bibnamefont {Sengstock}}, \bibinfo {author}
  {\bibfnamefont {V.}~\bibnamefont {Tamma}}, \bibinfo {author} {\bibfnamefont
  {T.}~\bibnamefont {Valenzuela}}, \bibinfo {author} {\bibfnamefont
  {A.}~\bibnamefont {Vogel}}, \bibinfo {author} {\bibfnamefont
  {R.}~\bibnamefont {Walser}}, \bibinfo {author} {\bibfnamefont
  {T.}~\bibnamefont {Wendrich}}, \bibinfo {author} {\bibfnamefont
  {P.}~\bibnamefont {Windpassinger}}, \bibinfo {author} {\bibfnamefont
  {W.}~\bibnamefont {Zeller}}, \bibinfo {author} {\bibfnamefont
  {T.}~\bibnamefont {van Zoest}}, \bibinfo {author} {\bibfnamefont
  {W.}~\bibnamefont {Ertmer}}, \bibinfo {author} {\bibfnamefont {W.~P.}\
  \bibnamefont {Schleich}}, \ and\ \bibinfo {author} {\bibfnamefont {E.~M.}\
  \bibnamefont {Rasel}},\ }\href {\doibase 10.1103/PhysRevLett.110.093602}
  {\bibfield  {journal} {\bibinfo  {journal} {Phys. Rev. Lett.}\ }\textbf
  {\bibinfo {volume} {110}},\ \bibinfo {pages} {093602} (\bibinfo {year}
  {2013})}\BibitemShut {NoStop}%
\bibitem [{\citenamefont {Sugarbaker}\ \emph {et~al.}(2013)\citenamefont
  {Sugarbaker}, \citenamefont {Dickerson}, \citenamefont {Hogan}, \citenamefont
  {Johnson},\ and\ \citenamefont {Kasevich}}]{sugarbaker13}%
  \BibitemOpen
  \bibfield  {author} {\bibinfo {author} {\bibfnamefont {A.}~\bibnamefont
  {Sugarbaker}}, \bibinfo {author} {\bibfnamefont {S.~M.}\ \bibnamefont
  {Dickerson}}, \bibinfo {author} {\bibfnamefont {J.~M.}\ \bibnamefont
  {Hogan}}, \bibinfo {author} {\bibfnamefont {D.~M.~S.}\ \bibnamefont
  {Johnson}}, \ and\ \bibinfo {author} {\bibfnamefont {M.~A.}\ \bibnamefont
  {Kasevich}},\ }\href {\doibase 10.1103/PhysRevLett.111.113002} {\bibfield
  {journal} {\bibinfo  {journal} {Phys. Rev. Lett.}\ }\textbf {\bibinfo
  {volume} {111}},\ \bibinfo {pages} {113002} (\bibinfo {year}
  {2013})}\BibitemShut {NoStop}%
\bibitem [{\citenamefont {L\'ev\'eque}\ \emph {et~al.}(2009)\citenamefont
  {L\'ev\'eque}, \citenamefont {Gauguet}, \citenamefont {Michaud},
  \citenamefont {Pereira Dos~Santos},\ and\ \citenamefont
  {Landragin}}]{leveque09}%
  \BibitemOpen
  \bibfield  {author} {\bibinfo {author} {\bibfnamefont {T.}~\bibnamefont
  {L\'ev\'eque}}, \bibinfo {author} {\bibfnamefont {A.}~\bibnamefont
  {Gauguet}}, \bibinfo {author} {\bibfnamefont {F.}~\bibnamefont {Michaud}},
  \bibinfo {author} {\bibfnamefont {F.}~\bibnamefont {Pereira Dos~Santos}}, \
  and\ \bibinfo {author} {\bibfnamefont {A.}~\bibnamefont {Landragin}},\ }\href
  {\doibase 10.1103/PhysRevLett.103.080405} {\bibfield  {journal} {\bibinfo
  {journal} {Phys. Rev. Lett.}\ }\textbf {\bibinfo {volume} {103}},\ \bibinfo
  {pages} {080405} (\bibinfo {year} {2009})}\BibitemShut {NoStop}%
\bibitem [{\citenamefont {Giese}\ \emph {et~al.}(2013)\citenamefont {Giese},
  \citenamefont {Roura}, \citenamefont {Tackmann}, \citenamefont {Rasel},\ and\
  \citenamefont {Schleich}}]{giese13}%
  \BibitemOpen
  \bibfield  {author} {\bibinfo {author} {\bibfnamefont {E.}~\bibnamefont
  {Giese}}, \bibinfo {author} {\bibfnamefont {A.}~\bibnamefont {Roura}},
  \bibinfo {author} {\bibfnamefont {G.}~\bibnamefont {Tackmann}}, \bibinfo
  {author} {\bibfnamefont {E.~M.}\ \bibnamefont {Rasel}}, \ and\ \bibinfo
  {author} {\bibfnamefont {W.~P.}\ \bibnamefont {Schleich}},\ }\href {\doibase
  10.1103/PhysRevA.88.053608} {\bibfield  {journal} {\bibinfo  {journal} {Phys.
  Rev. A}\ }\textbf {\bibinfo {volume} {88}},\ \bibinfo {pages} {053608}
  (\bibinfo {year} {2013})}\BibitemShut {NoStop}%
\bibitem [{\citenamefont {Malossi}\ \emph {et~al.}(2010)\citenamefont
  {Malossi}, \citenamefont {Bodart}, \citenamefont {Merlet}, \citenamefont
  {Lévèque}, \citenamefont {Landragin},\ and\ \citenamefont {Pereira
  Dos~Santos}}]{malossi10}%
  \BibitemOpen
  \bibfield  {author} {\bibinfo {author} {\bibfnamefont {N.}~\bibnamefont
  {Malossi}}, \bibinfo {author} {\bibfnamefont {Q.}~\bibnamefont {Bodart}},
  \bibinfo {author} {\bibfnamefont {S.}~\bibnamefont {Merlet}}, \bibinfo
  {author} {\bibfnamefont {T.}~\bibnamefont {Lévèque}}, \bibinfo {author}
  {\bibfnamefont {A.}~\bibnamefont {Landragin}}, \ and\ \bibinfo {author}
  {\bibfnamefont {F.}~\bibnamefont {Pereira Dos~Santos}},\ }\href {\doibase
  10.1103/PhysRevA.81.013617} {\bibfield  {journal} {\bibinfo  {journal} {Phys.
  Rev. A}\ }\textbf {\bibinfo {volume} {81}},\ \bibinfo {pages} {013617}
  (\bibinfo {year} {2010})}\BibitemShut {NoStop}%
\bibitem [{\citenamefont {Lan}\ \emph {et~al.}(2012)\citenamefont {Lan},
  \citenamefont {Kuan}, \citenamefont {Estey}, \citenamefont {Haslinger},\ and\
  \citenamefont {M\"{u}ller}}]{lan12}%
  \BibitemOpen
  \bibfield  {author} {\bibinfo {author} {\bibfnamefont {S.-Y.}\ \bibnamefont
  {Lan}}, \bibinfo {author} {\bibfnamefont {P.-C.}\ \bibnamefont {Kuan}},
  \bibinfo {author} {\bibfnamefont {B.}~\bibnamefont {Estey}}, \bibinfo
  {author} {\bibfnamefont {P.}~\bibnamefont {Haslinger}}, \ and\ \bibinfo
  {author} {\bibfnamefont {H.}~\bibnamefont {M\"{u}ller}},\ }\href {\doibase
  10.1103/PhysRevLett.108.090402} {\bibfield  {journal} {\bibinfo  {journal}
  {Phys. Rev. Lett.}\ }\textbf {\bibinfo {volume} {108}},\ \bibinfo {pages}
  {090402} (\bibinfo {year} {2012})}\BibitemShut {NoStop}%
\bibitem [{\citenamefont {Dickerson}\ \emph {et~al.}(2013)\citenamefont
  {Dickerson}, \citenamefont {Hogan}, \citenamefont {Sugarbaker}, \citenamefont
  {Johnson},\ and\ \citenamefont {Kasevich}}]{dickerson13}%
  \BibitemOpen
  \bibfield  {author} {\bibinfo {author} {\bibfnamefont {S.~M.}\ \bibnamefont
  {Dickerson}}, \bibinfo {author} {\bibfnamefont {J.~M.}\ \bibnamefont
  {Hogan}}, \bibinfo {author} {\bibfnamefont {A.}~\bibnamefont {Sugarbaker}},
  \bibinfo {author} {\bibfnamefont {D.~M.~S.}\ \bibnamefont {Johnson}}, \ and\
  \bibinfo {author} {\bibfnamefont {M.~A.}\ \bibnamefont {Kasevich}},\ }\href
  {\doibase 10.1103/PhysRevLett.111.083001} {\bibfield  {journal} {\bibinfo
  {journal} {Phys. Rev. Lett.}\ }\textbf {\bibinfo {volume} {111}},\ \bibinfo
  {pages} {083001} (\bibinfo {year} {2013})}\BibitemShut {NoStop}%
\bibitem [{\citenamefont {Snadden}\ \emph {et~al.}(1998)\citenamefont
  {Snadden}, \citenamefont {McGuirk}, \citenamefont {Bouyer}, \citenamefont
  {Haritos},\ and\ \citenamefont {Kasevich}}]{snadden98}%
  \BibitemOpen
  \bibfield  {author} {\bibinfo {author} {\bibfnamefont {M.~J.}\ \bibnamefont
  {Snadden}}, \bibinfo {author} {\bibfnamefont {J.~M.}\ \bibnamefont
  {McGuirk}}, \bibinfo {author} {\bibfnamefont {P.}~\bibnamefont {Bouyer}},
  \bibinfo {author} {\bibfnamefont {K.~G.}\ \bibnamefont {Haritos}}, \ and\
  \bibinfo {author} {\bibfnamefont {M.~A.}\ \bibnamefont {Kasevich}},\ }\href
  {\doibase 10.1103/PhysRevLett.81.971} {\bibfield  {journal} {\bibinfo
  {journal} {Phys. Rev. Lett.}\ }\textbf {\bibinfo {volume} {81}},\ \bibinfo
  {pages} {971} (\bibinfo {year} {1998})}\BibitemShut {NoStop}%
\bibitem [{\citenamefont {Rosi}\ \emph {et~al.}(2015)\citenamefont {Rosi},
  \citenamefont {Cacciapuoti}, \citenamefont {Sorrentino}, \citenamefont
  {Menchetti}, \citenamefont {Prevedelli},\ and\ \citenamefont
  {Tino}}]{rosi15}%
  \BibitemOpen
  \bibfield  {author} {\bibinfo {author} {\bibfnamefont {G.}~\bibnamefont
  {Rosi}}, \bibinfo {author} {\bibfnamefont {L.}~\bibnamefont {Cacciapuoti}},
  \bibinfo {author} {\bibfnamefont {F.}~\bibnamefont {Sorrentino}}, \bibinfo
  {author} {\bibfnamefont {M.}~\bibnamefont {Menchetti}}, \bibinfo {author}
  {\bibfnamefont {M.}~\bibnamefont {Prevedelli}}, \ and\ \bibinfo {author}
  {\bibfnamefont {G.~M.}\ \bibnamefont {Tino}},\ }\href {\doibase
  10.1103/PhysRevLett.114.013001} {\bibfield  {journal} {\bibinfo  {journal}
  {Phys. Rev. Lett.}\ }\textbf {\bibinfo {volume} {114}},\ \bibinfo {pages}
  {013001} (\bibinfo {year} {2015})}\BibitemShut {NoStop}%
\bibitem [{\citenamefont {{Geiger}}\ \emph {et~al.}()\citenamefont {{Geiger}},
  \citenamefont {{Amand}}, \citenamefont {{Bertoldi}}, \citenamefont
  {{Canuel}}, \citenamefont {{Chaibi}}, \citenamefont {{Danquigny}},
  \citenamefont {{Dutta}}, \citenamefont {{Fang}}, \citenamefont {{Gaffet}},
  \citenamefont {{Gillot}}, \citenamefont {{Holleville}}, \citenamefont
  {{Landragin}}, \citenamefont {{Merzougui}}, \citenamefont {{Riou}},
  \citenamefont {{Savoie}},\ and\ \citenamefont {{Bouyer}}}]{geiger15}%
  \BibitemOpen
  \bibfield  {author} {\bibinfo {author} {\bibfnamefont {R.}~\bibnamefont
  {{Geiger}}}, \bibinfo {author} {\bibfnamefont {L.}~\bibnamefont {{Amand}}},
  \bibinfo {author} {\bibfnamefont {A.}~\bibnamefont {{Bertoldi}}}, \bibinfo
  {author} {\bibfnamefont {B.}~\bibnamefont {{Canuel}}}, \bibinfo {author}
  {\bibfnamefont {W.}~\bibnamefont {{Chaibi}}}, \bibinfo {author}
  {\bibfnamefont {C.}~\bibnamefont {{Danquigny}}}, \bibinfo {author}
  {\bibfnamefont {I.}~\bibnamefont {{Dutta}}}, \bibinfo {author} {\bibfnamefont
  {B.}~\bibnamefont {{Fang}}}, \bibinfo {author} {\bibfnamefont
  {S.}~\bibnamefont {{Gaffet}}}, \bibinfo {author} {\bibfnamefont
  {J.}~\bibnamefont {{Gillot}}}, \bibinfo {author} {\bibfnamefont
  {D.}~\bibnamefont {{Holleville}}}, \bibinfo {author} {\bibfnamefont
  {A.}~\bibnamefont {{Landragin}}}, \bibinfo {author} {\bibfnamefont
  {M.}~\bibnamefont {{Merzougui}}}, \bibinfo {author} {\bibfnamefont
  {I.}~\bibnamefont {{Riou}}}, \bibinfo {author} {\bibfnamefont
  {D.}~\bibnamefont {{Savoie}}}, \ and\ \bibinfo {author} {\bibfnamefont
  {P.}~\bibnamefont {{Bouyer}}},\ }\href@noop {} {\bibinfo  {journal}
  {\texttt{arXiv:1505.07137}}\ }\BibitemShut {NoStop}%
\bibitem [{\citenamefont {Dimopoulos}\ \emph {et~al.}(2008)\citenamefont
  {Dimopoulos}, \citenamefont {Graham}, \citenamefont {Hogan}, \citenamefont
  {Kasevich},\ and\ \citenamefont {Rajendran}}]{dimopoulos08b}%
  \BibitemOpen
\bibfield  {journal} {  }\bibfield  {author} {\bibinfo {author} {\bibfnamefont
  {S.}~\bibnamefont {Dimopoulos}}, \bibinfo {author} {\bibfnamefont {P.~W.}\
  \bibnamefont {Graham}}, \bibinfo {author} {\bibfnamefont {J.~M.}\
  \bibnamefont {Hogan}}, \bibinfo {author} {\bibfnamefont {M.~A.}\ \bibnamefont
  {Kasevich}}, \ and\ \bibinfo {author} {\bibfnamefont {S.}~\bibnamefont
  {Rajendran}},\ }\href {\doibase 10.1103/PhysRevD.78.122002} {\bibfield
  {journal} {\bibinfo  {journal} {Phys. Rev. D}\ }\textbf {\bibinfo {volume}
  {78}},\ \bibinfo {pages} {122002} (\bibinfo {year} {2008})}\BibitemShut
  {NoStop}%
\bibitem [{\citenamefont {Graham}\ \emph {et~al.}(2013)\citenamefont {Graham},
  \citenamefont {Hogan}, \citenamefont {Kasevich},\ and\ \citenamefont
  {Rajendran}}]{graham13}%
  \BibitemOpen
  \bibfield  {author} {\bibinfo {author} {\bibfnamefont {P.~W.}\ \bibnamefont
  {Graham}}, \bibinfo {author} {\bibfnamefont {J.~M.}\ \bibnamefont {Hogan}},
  \bibinfo {author} {\bibfnamefont {M.~A.}\ \bibnamefont {Kasevich}}, \ and\
  \bibinfo {author} {\bibfnamefont {S.}~\bibnamefont {Rajendran}},\ }\href
  {\doibase 10.1103/PhysRevLett.110.171102} {\bibfield  {journal} {\bibinfo
  {journal} {Phys. Rev. Lett.}\ }\textbf {\bibinfo {volume} {110}},\ \bibinfo
  {pages} {171102} (\bibinfo {year} {2013})}\BibitemShut {NoStop}%
\bibitem [{\citenamefont {Hogan}\ and\ \citenamefont {Kasevich}()}]{hogan15}%
  \BibitemOpen
  \bibfield  {author} {\bibinfo {author} {\bibfnamefont {J.~M.}\ \bibnamefont
  {Hogan}}\ and\ \bibinfo {author} {\bibfnamefont {M.~A.}\ \bibnamefont
  {Kasevich}},\ }\href@noop {} {\bibinfo  {journal}
  {\texttt{arXiv:1501.06797}}\ }\BibitemShut {NoStop}%
\bibitem [{\citenamefont {Antoine}\ and\ \citenamefont
  {Bord\'{e}}(2003)}]{antoine03b}%
  \BibitemOpen
\bibfield  {journal} {  }\bibfield  {author} {\bibinfo {author} {\bibfnamefont
  {C.}~\bibnamefont {Antoine}}\ and\ \bibinfo {author} {\bibfnamefont {C.~J.}\
  \bibnamefont {Bord\'{e}}},\ }\href
  {http://stacks.iop.org/1464-4266/5/i=2/a=380} {\bibfield  {journal} {\bibinfo
   {journal} {J. Opt. B: Quantum and Semiclass. Opt.}\ }\textbf {\bibinfo
  {volume} {5}},\ \bibinfo {pages} {S199} (\bibinfo {year} {2003})}\BibitemShut
  {NoStop}%
\bibitem [{\citenamefont {Kleinert}\ \emph {et~al.}(2015)\citenamefont
  {Kleinert}, \citenamefont {Kajari}, \citenamefont {Roura},\ and\
  \citenamefont {Schleich}}]{kleinert15}%
  \BibitemOpen
  \bibfield  {author} {\bibinfo {author} {\bibfnamefont {S.}~\bibnamefont
  {Kleinert}}, \bibinfo {author} {\bibfnamefont {E.}~\bibnamefont {Kajari}},
  \bibinfo {author} {\bibfnamefont {A.}~\bibnamefont {Roura}}, \ and\ \bibinfo
  {author} {\bibfnamefont {W.~P.}\ \bibnamefont {Schleich}},\ }\href@noop {}
  {\bibfield  {journal} {\bibinfo  {journal} {Physics Reports}\ } (\bibinfo
  {year} {2015})},\ \bibinfo {note} {accepted for publication}\BibitemShut
  {NoStop}%
\bibitem [{\citenamefont {Schleich}(2001)}]{schleich}%
  \BibitemOpen
  \bibfield  {author} {\bibinfo {author} {\bibfnamefont {W.~P.}\ \bibnamefont
  {Schleich}},\ }\href@noop {} {\emph {\bibinfo {title} {Quantum Optics in
  Phase Space}}}\ (\bibinfo  {publisher} {Wiley-VCH},\ \bibinfo {year}
  {2001})\BibitemShut {NoStop}%
\end{thebibliography}%
